\PassOptionsToPackage{dvipsnames}{xcolor}

\documentclass[sn-mathphys-num]{sn-jnl}

\usepackage{mathtools}
\usepackage{mathrsfs} 
\usepackage{pifont}
\usepackage{tikz-cd}
\usepackage{enumitem}
\usepackage{amssymb}
\usepackage[stdtext,vcentermath]{myyoungtab}
\usepackage{bm} 
\usepackage[section]{placeins}
\usetikzlibrary{positioning}
\theoremstyle{definition}
\newtheorem{example}{Example}
\newcommand{\circnum}[1]{\ding{\numexpr171+#1\relax}}
\newcommand{\mybox}[1]{\makebox[2.0em]{$#1$}} 
\newcommand{\myboxx}[1]{\makebox[3.0em]{$#1$}}
\newcommand{\myboxxx}[1]{\makebox[4.0em]{$#1$}}

\newcommand{\blue}[1]{\textcolor{blue}{#1}}
\newcommand{\red}[1]{\textcolor{red}{#1}}
\newcommand{\green}[1]{\textcolor{green!70!black}{#1}}
\newcommand{\black}[1]{\textcolor{black}{#1}}
\newcommand\et[1]{\tilde{e}_{#1}}
\newcommand\ft[1]{\tilde{f}_{#1}}

\title[]{Novel energy preserving bijections between
affine crystals for $U_q(\widehat{\mathfrak{sl}}_2)$ and integer partitions}

\author{\fnm{Sota} \sur{Miyazawa}}\email{sota.myzw1@gmail.com}
\equalcont{These authors contributed equally to this work.}

\author{\fnm{Taichiro} \sur{Takagi}}\email{takagi@nda.ac.jp}
\equalcont{These authors contributed equally to this work.}

\affil{\orgdiv{Department of Applied Physics}, \orgname{National Defense Academy}, \orgaddress{\state{Kanagawa 239-8686}, \country{Japan}}}

\abstract{
Let $B(\Lambda_a) \, (a=0,1)$ be the crystal of the level 1 integrable irreducible highest weight representation of the affine quantum group $U_q(\widehat{\mathfrak{sl}}_2)$. We consider the crystal graphs of degree $n$ associated with the irreducible $(2r+1)$-dimensional (resp. $(2r+2)$-dimensional) $U_q(\mathfrak{sl}_2)$ module in $B(\Lambda_0)$ (resp. $B(\Lambda_1)$). In this paper, we construct an explicit combinatorial procedure providing a bijection between the set of highest weight paths in these graphs with respect to the action of the Kashiwara operator $\ft{1}$, and the set of integer partitions of $n$ with sqrank (resp. rerank) $r$, {which is a recently introduced partition statistic.
As a byproduct, we also obtain a precise interpretation of the motif description of spinons 
suggested by 
Bernard-Pasquier-Serban in the spinon picture for
Wess-Zumino-Witten conformal field theory models.}}

\keywords{affine crystals, integer partitions}

\begin{document}

\maketitle

\begin{abstract}

\end{abstract}

\section{Introduction}\label{sec1}
Recently, the second author introduced a pair of new partition statistics called
sqrank and rerank \cite{Takagi2026}. 
It was motivated by  
the study of the minimal excludant in integer partitions
by G.E. Andrews and D. Newman \cite{AN2020}.
For all nonnegative integers $n$, it was shown that
the partitions of $n$ on which sqrank or rerank takes on a particular value, say $r$, are equinumerous with the partitions of $n$ on which the
odd/even minimal {excludant} takes on the corresponding value, $2r+1$ or $2r+2$.

{In this paper, we explore a study for clarifying the meaning of the above partition statistics more
deeply, and seek to potential applications to problems in mathematical physics.}
It is based on {the} observation presented in {\cite[Appendix B]{Takagi2026}},
which states that the generating functions of the number of the
partitions with specified values of such statistics are
related to characters of affine Lie algebra $\widehat{\mathfrak{sl}}_2$.
Let $V(\Lambda_0)$ and $V(\Lambda_1)$ be the level $1$ integrable
irreducible highest weight modules of
the quantum affine Lie algebra 
$U_q(\widehat{\mathfrak{sl}}_2)$ with highest weights $\Lambda_0$ and $\Lambda_1$.
By regarding them as modules for a subalgebra $\mathfrak{sl}_2 \subset \widehat{\mathfrak{sl}}_2$,
they can be decomposed into irreducible finite dimensional $U_q(\mathfrak{sl}_2)$ modules.
From Weyl-Kac character formula of the level $1$ integrable modules of $\widehat{\mathfrak{sl}}_2$ 
\cite{Kac1990, KacRaina1987}, one can derive the following expressions for 
the relevant branching functions:
\begin{equation}\label{branchingFunction}
    \begin{aligned}
        b^{V(\Lambda_0)}_{2r \overline{\Lambda}_1}(q) &=\frac{q^{r^2}(1 - q^{2r+1})}{(q;q)_{\infty}} =
        \frac{q^{1 + 3 + \cdots + (2 r-1)}}{\prod_{m=1, m \ne 2r+1}^\infty (1-q^m)},\\
        b^{V(\Lambda_1)}_{(2r+1) \overline{\Lambda}_1}(q) &=\frac{q^{r(r+1)}(1 - q^{2(r+1)})}{(q;q)_{\infty}} =
        \frac{q^{2 + 4 + \cdots + 2 r}}{\prod_{m=1, m \ne 2r+2}^\infty (1-q^m)}.
    \end{aligned}
\end{equation}
{Here, 
the subscript $m \overline{\Lambda}_1 \, (m=2r, 2r+1)$ indicates
that it is for the $m+1$ dimensional irreducible module for the subalgebra $\mathfrak{sl}_2$.}
By these expressions one can deduce the following statements.
For each {\em degree} $n$ measuring the depth along the direction of the null root $\delta$
in the affine weight space of $\widehat{\mathfrak{sl}}_2$,
the number of $2r+1$ dimensional $U_q(\mathfrak{sl}_2)$ modules in
$V(\Lambda_0)$ is
equal to the number of partitions of $n$ with odd minimal excludant $2r+1$, and that of $2r+2$ dimensional $U_q(\mathfrak{sl}_2)$ modules in $V(\Lambda_1)$ is equal to the number of partitions of $n$ with even minimal excludant $2r+2$.
{Motivated} by this observation along with {the above mentioned equinumerous property} of
our new partition statistics,
we try to find an explicit combinatorial procedure to describe
energy-preserving bijections between
certain representation theoretical objects, known as the {\em crystal graphs}
in the theory of crystal bases \cite{Kashiwara1991, Kashiwara1994, HK2002, BS2017}, and the 
set of integer partitions \cite{Andrews1984}.

In order to describe the above decomposition explicitly,
we shall use the Kyoto path model \cite{BS2017} of the
affine crystals $B(\Lambda_i)$ for $V(\Lambda_i) \, (i=0,1)$ developed 
by S.-J.~Kang et al.~\cite{KMN}.
The crystals are represented by colored oriented graphs, known as crystal graphs.
Let us show an example that is relevant to the present work.
\begin{equation}\label{t:cry:b1b2}
\setlength{\unitlength}{0.35mm}
\begin{picture}(240,30)(-7,0)
\put(-15,10){$B:$}
\put(20,10){$\young(0) \qquad \quad\young(1)$} \put(35,9){\vector(1,0){20}}\put(55,15){\vector(-1,0){20}}
\put(43,19){${\scriptstyle 0}$}\put(43,-1){${\scriptstyle 1}$}
\end{picture}
\end{equation}
Here the arrows with index $i$ represent the actions of the Kashiwara operator $\tilde{f}_i$.
This is the crystal for a two-dimensional irreducible representation of $U_q'(\widehat{\mathfrak{sl}}_2)$,
which is known as a classical crystal with null root $\delta = 0$.
Since it has a property known as level $1$ perfectness, 
the following isomorphisms of crystals hold \cite{KMN}.
\begin{equation*}
B(\Lambda_0) \otimes B \cong B(\Lambda_1), \qquad
B(\Lambda_1) \otimes B \cong B(\Lambda_0).
\end{equation*}
By using these {isomorphisms} repeatedly, 
the elements of these affine crystals can be explicitly realized as semi-{infinite} bit sequences
called {\em paths}, and  
the highest weight elements of
$B(\Lambda_0)$ and $B(\Lambda_1)$ are given by
\begin{equation*}
\bar{p}^{\Lambda_0} = \dots \otimes \young(0) \otimes \young(1) \otimes \young(0) \otimes
\young(1), \quad \mbox{and} \quad 
\bar{p}^{\Lambda_1} = \dots \otimes \young(1) \otimes \young(0) \otimes \young(1) \otimes
\young(0), 
\end{equation*}
respectively.
In what follows, we simply write them as $\bar{p}^{\Lambda_0} = \cdots 0101$ and 
$\bar{p}^{\Lambda_1} = \cdots 1010$.
Borrowing a terminology from physics,
we call them the 
{\em ground states}.

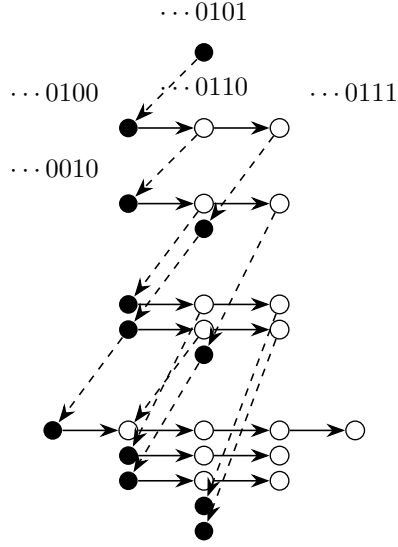
\begin{figure}[hbtp]
\centering
\begin{tikzpicture}[edge/.style={-{Stealth[length=2.5mm,width=1.8mm]}, line width=0.6pt}]
  \node[circle, fill=black, inner sep=0pt, minimum size=7pt] (top) at (0,0) {};
  \node[circle, fill=black, inner sep=0pt, minimum size=7pt] (left) at (-1,-1) {};
  \node[circle, draw=black, fill=white, line width=0.5pt, inner sep=0pt, minimum size=7pt] (middle) at (0,-1) {};
  \node[circle, draw=black, fill=white, line width=0.5pt, inner sep=0pt, minimum size=7pt] (right) at (1,-1) {};
  \node[circle, fill=black, inner sep=0pt, minimum size=7pt] (lowerleft) at (-1,-2) {};
  \node[circle, draw=black, fill=white, line width=0.5pt, inner sep=0pt, minimum size=7pt] (lowermiddle) at (0,-2) {};
  \node[circle, draw=black, fill=white, line width=0.5pt, inner sep=0pt, minimum size=7pt] (lowerright) at (1,-2) {};
  \node[circle, fill=black, inner sep=0pt, minimum size=7pt] (bottom) at (0,-2.333) {};
  \node[circle, fill=black, inner sep=0pt, minimum size=7pt] (rowfiveleft) at (-1,-3.333) {};
  \node[circle, draw=black, fill=white, line width=0.5pt, inner sep=0pt, minimum size=7pt] (rowfivemiddle) at (0,-3.333) {};
  \node[circle, draw=black, fill=white, line width=0.5pt, inner sep=0pt, minimum size=7pt] (rowfiveright) at (1,-3.333) {};
  \node[circle, fill=black, inner sep=0pt, minimum size=7pt] (rowsixleft) at (-1,-3.666) {};
  \node[circle, draw=black, fill=white, line width=0.5pt, inner sep=0pt, minimum size=7pt] (rowsixmiddle) at (0,-3.666) {};
  \node[circle, draw=black, fill=white, line width=0.5pt, inner sep=0pt, minimum size=7pt] (rowsixright) at (1,-3.666) {};
  \node[circle, fill=black, inner sep=0pt, minimum size=7pt] (rowsevenmiddle) at (0,-3.999) {};
  \node[circle, fill=black, inner sep=0pt, minimum size=7pt] (roweightlefttwo) at (-2,-4.999) {};
  \node[circle, draw=black, fill=white, line width=0.5pt, inner sep=0pt, minimum size=7pt] (roweightleftone) at (-1,-4.999) {};
  \node[circle, draw=black, fill=white, line width=0.5pt, inner sep=0pt, minimum size=7pt] (roweightmiddle) at (0,-4.999) {};
  \node[circle, draw=black, fill=white, line width=0.5pt, inner sep=0pt, minimum size=7pt] (roweightrightone) at (1,-4.999) {};
  \node[circle, draw=black, fill=white, line width=0.5pt, inner sep=0pt, minimum size=7pt] (roweightrighttwo) at (2,-4.999) {};
  \node[circle, fill=black, inner sep=0pt, minimum size=7pt] (rownineleft) at (-1,-5.332) {};
  \node[circle, draw=black, fill=white, line width=0.5pt, inner sep=0pt, minimum size=7pt] (rowninemiddle) at (0,-5.332) {};
  \node[circle, draw=black, fill=white, line width=0.5pt, inner sep=0pt, minimum size=7pt] (rownineright) at (1,-5.332) {};
  \node[circle, fill=black, inner sep=0pt, minimum size=7pt] (rowtenleft) at (-1,-5.665) {};
  \node[circle, draw=black, fill=white, line width=0.5pt, inner sep=0pt, minimum size=7pt] (rowtenmiddle) at (0,-5.665) {};
  \node[circle, draw=black, fill=white, line width=0.5pt, inner sep=0pt, minimum size=7pt] (rowtenright) at (1,-5.665) {};
  \node[circle, fill=black, inner sep=0pt, minimum size=7pt] (rowelevenmiddle) at (0,-5.998) {};
  \node[circle, fill=black, inner sep=0pt, minimum size=7pt] (rowtwelvemiddle) at (0,-6.331) {};
  \node[above=5pt of top] {$\cdots0101$};
  \node[above left=4pt and 5pt of left] {$\cdots0100$};
  \node[above=5pt of middle] {$\cdots0110$};
  \node[above right=4pt and 5pt of right] {$\cdots0111$};
  \node[above left=4pt and 5pt of lowerleft] {$\cdots0010$};
  \draw[edge, dashed] (top) -- (left);
  \draw[edge, dashed] (middle) -- (lowerleft);
  \draw[edge, dashed] (right) -- (bottom);
  \draw[edge, dashed] (lowermiddle) -- (rowfiveleft);
  \draw[edge, dashed] (bottom) -- (rowsixleft);
  \draw[edge, dashed] (lowerright) -- (rowsevenmiddle);
  \draw[edge, dashed] (rowfivemiddle) -- (rownineleft);
  \draw[edge, dashed] (rowfiveright) -- (rowelevenmiddle);
  \draw[edge, dashed] (rowsixleft) -- (roweightlefttwo);
  \draw[edge, dashed] (rowsixmiddle) -- (roweightleftone);
  \draw[edge, dashed] (rowsixright) -- (rowtwelvemiddle);
  \draw[edge, dashed] (rowsevenmiddle) -- (rowtenleft);
  \draw[edge] (left) -- (middle);
  \draw[edge] (middle) -- (right);
  \draw[edge] (lowerleft) -- (lowermiddle);
  \draw[edge] (lowermiddle) -- (lowerright);
  \draw[edge] (rowfiveleft) -- (rowfivemiddle);
  \draw[edge] (rowfivemiddle) -- (rowfiveright);
  \draw[edge] (rowsixleft) -- (rowsixmiddle);
  \draw[edge] (rowsixmiddle) -- (rowsixright);
  \draw[edge] (roweightlefttwo) -- (roweightleftone);
  \draw[edge] (roweightleftone) -- (roweightmiddle);
  \draw[edge] (roweightmiddle) -- (roweightrightone);
  \draw[edge] (roweightrightone) -- (roweightrighttwo);
  \draw[edge] (rownineleft) -- (rowninemiddle);
  \draw[edge] (rowninemiddle) -- (rownineright);
  \draw[edge] (rowtenleft) -- (rowtenmiddle);
  \draw[edge] (rowtenmiddle) -- (rowtenright);
\end{tikzpicture}
\caption{The part of the crystal graph of $B(\Lambda_0)$ with degree 0 to 4. (black nodes: $\ft{1}$-highest paths, dashed arrow: $\ft{0}$, solid arrow: $\ft{1}$.)}
\label{fig:crystal-graph}
\end{figure}

For simplicity, we only consider $B(\Lambda_0)$ in the rest of the present section.
A general state of $B(\Lambda_0)$ is a bit sequence that coincides with
$\bar{p}^{\Lambda_0}$ sufficiently far to the left.
One can obtain every state of $B(\Lambda_0)$ from the ground state $\bar{p}^{\Lambda_0}$
by
applying a suitably chosen sequence of
Kashiwara operators $\ft{0}$ and $\ft{1}$, according to the well known procedure for the tensor 
products of crystals \cite{KMN}, which may be described as follows.
For the action of $\ft{1}$,
we assign a left bracket ``$($" to each 0, and a right bracket 
 ``$)$" to each 1.
Read brackets from left to right.
If there exists an unmatched left bracket, turn the corresponding 0 into 1.
If otherwise, delete the path.
In an analogous way, the action of $\ft{0}$ is given by simply
exchanging the roles of $0$ and $1$.
For instance, we have $\ft{0} (\cdots 0101) = \cdots 0100$ and
$\ft{1} \ft{0}(\cdots 0101) = \cdots 0110$.
As a result, $B(\Lambda_0)$ is represented by a connected graph that
consists of infinite number of nodes and arrows,
the former correspond to the paths and the latter to the actions of $\ft{0}$ and $\ft{1}$. See Figure \ref{fig:crystal-graph}. 

Every node of $B(\Lambda_0)$ has its affine {\em weight}
$\Lambda_0 + r \alpha_1 -n \delta$, where $\alpha_1$ is a simple root, $\delta (\ne 0)$ is the
null root, and $r, n$ are integers.
We call $n$ the degree of the weight.
In particular
the ground state $\bar{p}^{\Lambda_0}$ has weight $\Lambda_0$,
which is the only element of degree 0.
The action of $\ft{0}$ increases the depth along the direction of the null root 
$\delta$, which amounts to adding 1 to the degree.
By ignoring all the arrows for $\ft{0}$, the crystal graph $B(\Lambda_0)$ decomposes into {an}
infinite number of connected {components, each of which consists of a finite number of}
the nodes for paths and the arrows for $\ft{1}$.
They are the crystal graphs for the crystal bases of the
irreducible finite dimensional $U_q(\mathfrak{sl}_2)$ modules.

As an example,
we illustrate all of 
the five crystal graphs for the 
$U_q(\mathfrak{sl}_2)$ modules with degree {four} in $B(\Lambda_0)$.
\begin{align*}
\cdots\mathbf{0000} &\longrightarrow & \cdots 1000 &\longrightarrow & \cdots 1100 &\longrightarrow & \cdots 1110 &\longrightarrow & \cdots1111, \\
&& \cdots \mathbf{00101010} &\longrightarrow & \cdots 10101010 &\longrightarrow & \cdots 10101011, && \\
&& \cdots \mathbf{001001} &\longrightarrow & \cdots 101001 &\longrightarrow & \cdots 101101, && \\
&&&& \cdots \mathbf{00101011}, &&&& \\
&&&& \cdots \mathbf{001101}. &&&&
\end{align*}
Here 
we let ``$\cdots$" denote {the} semi-infinite repetition of $01$s towards the {left, which can be identified with the ground state $\bar{p}^{\Lambda_0}$.}
The degree $n$ of a path $p=\cdots p_2 p_1$ can be written
as $n = E_{\leftarrow}^{\Lambda_0}(p)$, which is a sort of excitation energy of $p$
with respect to the ground state 
$\bar{p}^{\Lambda_0} = \cdots \bar{p}_4 \bar{p}_3 \bar{p}_2 \bar{p}_1 =\cdots 0101$ 
given by the formula 
\begin{equation}\label{eq:leftenergy}
E_{\leftarrow}^{\Lambda_0}(p) = \sum_{k=1}^\infty k \{ H_{\leftarrow}(p_{k+1},p_k) - H_{\leftarrow}(\bar{p}_{k+1},\bar{p}_k)\},
\end{equation}
where the {\em energy function} $H_{\leftarrow}(a,b)=1 (a \geq b), \, 0 (a < b)$
in the theory of affine crystals \cite{KMN} is used.
The five highest weight paths with respect to the action of $\ft{1}$ are written in bold font,
which we call {\em $\ft{1}$-highest paths} (h.p.'s).
There is one graph of {five} nodes (h.p. $\cdots\mathbf{0000}$), 
two graphs of {three} nodes (h.p.'s $\cdots\mathbf{00101010}$ and $\cdots\mathbf{001001}$), and two
graphs of {one} node (h.p.'s $\cdots\mathbf{00101011}$ and $\cdots\mathbf{001101}$).
We show that
there exists a natural combinatorial 
bijection between the set of partitions of $n$ with sqrank $r$ and
the set of $\ft{1}$-highest paths of the
degree $n$ crystal graphs 
for the irreducible $2r+1$-dimensional 
$U_q(\mathfrak{sl}_2)$ modules
in $B(\Lambda_0)$.
For every integer partition $\lambda$, we let $p(\lambda; \Lambda_0)$ denote
the  $\ft{1}$-highest path given by {this} bijection.
For the above mentioned five partitions of integer {four},
we shall show that our bijection yields the following result:
\begin{align*}
p((2,2); \Lambda_0) =& \cdots \mathbf{0000}, & & \\
p((1,1,1,1); \Lambda_0) =& \cdots \mathbf{00101010} ,&
p((3,1); \Lambda_0) =& \cdots \mathbf{001001} , \\
p((2,1,1); \Lambda_0) =& \cdots \mathbf{00101011} , &
p((4); \Lambda_0) =& \cdots\mathbf{ 001101} .
\end{align*}

The partition statistic sqrank 
was introduced in {\cite[Definition 1]{Takagi2026}.}
Let $\lambda$ be an integer partition, which can be regarded as a {Young} diagram.
To calculate its sqrank, first we decompose it into three sub-diagrams;
its Durfee square, its ``leg" located below, and its ``wing" located to the right.
From the top right end of its ``wing", we repeatedly strip off the rim hooks
of the same
arm length that equals
to the side of the Durfee square, say $n_0$, to get a residual diagram $R_0(\lambda)$.
Then a piece-wise linear function of the Frobenius representation of $R_0(\lambda)$
gives ${\rm sqrank}(\lambda)$, which takes an integer value between $0$ and $n_0$.
{See Sect.~\ref{subSect.partition} for more details.}

The method to construct the path $p(\lambda; \Lambda_0)$
from $\lambda$ by using our new bijection is described as follows.
We first get a bit sequence $p'_{R_0}(\lambda)$ of length $2n_0$ from the diagram
$R_0(\lambda)$ by using a
bijection introduced by the second author's former works \cite{Takagi2026, Takagi2025, Takagi2005}. 
This sequence can be viewed as an element of $B^{\otimes 2 n_0}$,
{where $B$ is the crystal in \eqref{t:cry:b1b2}.}
By applying Kashiwara operator $\et{1}$,
which may be viewed as an inverse of $\ft{1}$,
to $p'_{R_0}(\lambda)$ 
as many times as possible but not to delete it,
one obtains a $\ft{1}$-highest element $p_{R_0}(\lambda) \in B^{\otimes 2 n_0}$.
Now we replace every adjacent $01$ pair in $p_{R_0}(\lambda)$ with the sequence $0011$,
which amounts to getting a longer bit sequence $p_{R_0\uplus D_0}(\lambda)$.
By grouping every adjacent pair of bits into a block,
this bit sequence can be represented by a sequence of blocks.
We introduce the notion of {\em strings}, which are {particular types of contiguous
arrays of blocks classified into four types.}
We also introduce the notions of {\em 10-spots} and  {\em 01-spots},
which are {particular types of} spots between neighboring blocks where
the insertion of 10-blocks or 01-blocks is allowed.
One finds that there are exactly $n_0$ strings in $p_{R_0\uplus D_0}(\lambda)$, and each string possesses
exactly one 10-spot and one 01-spot.
Hence there are $2n_0$ such spots in total, and inserting a suitable 10- or 01-block into
the $k$-th spot amounts to increasing the energy $E_{\leftarrow}$ in 
\eqref{eq:leftenergy} by $k$.
Returning to the explanation of sqrank given above,
one finds that the ``leg" can contain parts of lengths ranging from $1$ to $n_0$, and
the rim hooks removed from the ``wing" can have lengths ranging from $n_0+1$ to $2 n_0$.
Now, for every such part or rim hook of length $k$ for $1 \leq k \leq 2n_0$, we shall insert a suitable 
10- or 01-block into the $k$-th spot of the sequence.
Then the final result yields the desired path $p(\lambda; \Lambda_0)$,
after {the} concatenation with the ground state {$\bar{p}^{\Lambda_0}$}.

The present work may be related to several models in mathematical physics.
In particular, 
the authors consider
it would be related to the spinon picture for
Wess-Zumino-Witten conformal field theory models
\cite{BPS1994, BLS1994}.
This comes from the following observation.
As one sees from 
{\cite[Sect.~8 and Appendices A/B]{Takagi2026}},
a natural expression for the generating function of the number of
partitions with specified values of sqrank or rerank coincides with
a special case of the spinon character formulas given by Hatayama et.~al.~ \cite{HKKOTY}.
Also, there was a study on the spinon picture and its related character formulas
based on the path model of affine crystals for $U_q(\widehat{\mathfrak{sl}}_2)$ \cite{NY1996}.
Although their precise relation is still unclear and needs further investigations,
{we shall propose a new interpretation of the motif description of spinons by Bernard et.~al.~\cite[Sect.~3]{BPS1994} in the present work.}

Throughout this paper, if we speak of {\em finite dimensional representations},
it means that those are not {representations} of $U_q(\widehat{\mathfrak{sl}}_2)$
but the irreducible representations of $U_q(\mathfrak{sl}_2)$, which 
{in their associated crystals}
can be
obtained by ignoring the actions of $\ft{0}$.

The remainder of this paper is organized as follows. 
In Sect.\ref{Sect.2}, we define an energy function and use it to define the energy for paths and finite bit strings. We also introduce integer partitions and their associated statistics here.
In Sects.\ref{Sect.R}-\ref{Sect.lambda}, we present a map that yields the corresponding crystal path for a given integer partition,
and in Sect.\ref{Sect.Reverse}, we verify {the existence of} its inverse map.
{Finally, we give a summary and discussions in Sect.~\ref{Sect.Summary_Discussions}.}

\section{Notations and Preliminaries}\label{Sect.2}
\subsection{Energy of a Path}
As we explained in Introduction, the elements of the crystals $B(\Lambda_0)$, $B(\Lambda_1)$ can be represented by semi-{infinite} bit sequences $p=\cdots p_2p_1$ indexed from right to left, which are called paths.
We define their {\em ground state}, the degrees of which are equal to zero, as
\begin{equation}\label{def:GroundState}
    \begin{aligned}
        \bar{p}^{\Lambda_0}=& \cdots \bar{p}_4^{\Lambda_0}\bar{p}_3^{\Lambda_0}\bar{p}_2^{\Lambda_0}\bar{p}_1^{\Lambda_0} = \cdots0101,\\
        \bar{p}^{\Lambda_1}=& \cdots \bar{p}_4^{\Lambda_1}\bar{p}_3^{\Lambda_1}\bar{p}_2^{\Lambda_1}\bar{p}_1^{\Lambda_1} = \cdots1010.
    \end{aligned}
\end{equation}
A general state of the crystal $B(\Lambda_0)$ is a bit sequence that coincides with $\bar{p}^{\Lambda_0}$ sufficiently far to the left; that is, recursively removing all adjacent 01-pairs leaves an even number of bits. On the other hand, a general state of the crystal $B(\Lambda_1)$ is a bit sequence that coincides with $\bar{p}^{\Lambda_1}$ sufficiently far to the left; that is, recursively removing all adjacent 01-pairs leaves an odd number of bits. 
If the bit sequences left contain no bit 1, then the original bit sequences are the highest states with respect to the action of the Kashiwara operator $\tilde{f}_1$. 

Using the {{\em energy functions}} defined by
\begin{equation}\label{def:EnergyFunction}
    \begin{aligned}
        H_{\to}(x,y)&\coloneq
        \begin{cases}
            1 & \text{$(x,y) = (0,1)$} \\
            0 & \text{$(x,y) \neq (0,1)$},
        \end{cases} \\
        H_{\gets}(x,y)&\coloneq 
        \begin{cases}
            0 & \text{$(x,y) = (0,1)$} \\
            1 & \text{$(x,y) \neq (0,1)$},
        \end{cases}
    \end{aligned}
\end{equation}
we define the {\em right energy} $E_{\to}(p)$ and the {\em left energy} $E_{\gets}^{\Lambda_a}(p)$ of a path $p=\cdots p_2p_1$ in the crystal $B(\Lambda_a)$ relative to the ground state $\bar{p}^{\Lambda_a}=\cdots\bar{p}_2^{\Lambda_a}\bar{p}_1^{\Lambda_a}$ as follows:
\begin{equation}\label{def:EnergyForPath}
    \begin{aligned}
        E_{\to}(p)&\coloneq \sum_{j=1}^{N-1}(N-j)H_{\to}(p_{j+1},p_j),\\
        E_{\gets}^{\Lambda_a}(p)&\coloneq\sum_{j=1}^{N-1} j[H_{\gets}(p_{j+1},p_j)-H_{\gets}(\bar{p}_{j+1}^{\Lambda_a},\bar{p}_j^{\Lambda_a})].
    \end{aligned}
\end{equation}
Here, $a \in \{0, 1\}$, and $N$ is a positive integer that is uniquely determined by the following conditions:
\begin{itemize}
    \item {$N\equiv a \pmod 2$}
    \item $p_{j}= \bar{p}_{j}^{\Lambda_a}\,\,(j>N) $
    \item $(p_N,p_{N-1})\neq(\bar{p}_N^{\Lambda_a},\bar{p}_{N-1}^{\Lambda_a})$ 
\end{itemize}
Throughout this paper, we identify the {finite} bit string $p_N\cdots p_2p_1$ with the {original} path $p=\cdots p_2p_1$. Furthermore, in general, for a sequence {$b=b_N\cdots b_2b_1$} of a finite number of bits, we define {its energies in an analogous way to the above case as};
\begin{equation}\label{def:EnergyForBitSequence}
    \begin{aligned}
        E_{\to}(b)&\coloneq \sum_{j=1}^{N-1}(N-j)H_{\to}(b_{j+1},b_j),\\
        E_{\gets}^{\Lambda_a}(b)&\coloneq\sum_{j=1}^{N-1} j[H_{\gets}(b_{j+1},b_j)-H_{\gets}(\bar{p}_{j+1}^{\Lambda_a},\bar{p}_j^{\Lambda_a})].
    \end{aligned}
\end{equation}
In addition, we {formally} set $p_0\coloneq0$ {as the 0-th bit of a path $p=...p_2 p_1$ in} this paper.

\subsection{Partitions of {an} Integer}\label{subSect.partition}
In this section, we introduce integer partitions, their corresponding Young diagrams, and the associated statistics used in this paper. 
A {\em partition} of a positive integer $n$ is a finite weakly decreasing sequence of positive integers $\lambda = (\lambda_1, \dots, \lambda_r)$ such that $\sum_{i=1}^r \lambda_i =n$, and each $\lambda_i$ is called a {\em part} of a partition $\lambda$. 
A {\em Young diagram} is a graphical representation of a partition, so we let a Young diagram corresponding a partition $\lambda$ to be also denoted by $\lambda$.
{In addition, we formally define a unique partition of zero and denote it as $\emptyset$.}
The element $(i,j)$, {which is called a} {\em cell}, of a Young diagram $\lambda$ is occupied by a unit square if $1\le j\le\lambda_i$ for $1\le i\le r$. 
We also use the {\em Frobenius representation} $F(\lambda) \coloneq (x_1, \dots, x_d \mid y_1, \dots, y_d)$. In this notation, $x_i = \#\{(i,j)\in\lambda \mid j>i\}$ represents the number of cells in the $i$-th row to the right of $(i,i)$, and $y_i = \#\{(j,i)\in\lambda \mid j>i\}$ represents the number of cells in the $i$-th column below $(i,i)$.
%
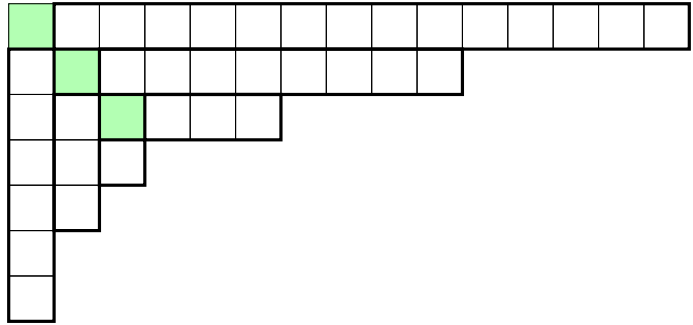
\begin{figure}[htbp]
  \centering
  \begin{tikzpicture}[x=6mm,y=6mm]
    \foreach \row/\len in {1/15,2/10,3/6,4/3,5/2,6/1,7/1} {
      \foreach \col in {1,...,\len} {
        \ifnum\row=\col
          \fill[green!30] (\col-1,-\row+1) rectangle (\col,-\row+2);
        \fi
        \draw (\col-1,-\row+1) rectangle (\col,-\row+2);
      }
    }
    \draw[very thick] (1,0) rectangle (15,1);
    \draw[very thick] (2,-1) rectangle (10,0);
    \draw[very thick] (3,-2) rectangle (6,-1);
    \draw[very thick] (0,-6) rectangle (1,0);
    \draw[very thick] (1,-4) rectangle (2,-1);
    \draw[very thick] (2,-3) rectangle (3,-2);
  \end{tikzpicture}
  \caption{Frobenius representation of $\lambda=(15,10,6,3,2,1,1)$; $F(\lambda)=(14,8,3\mid 6,3,1)$.}
\end{figure}
%
Suppose that the main diagonal of a Young diagram $\lambda$ consists of $d$ cells $(i,i)$ for $1 \le i \le d$. The square defined by this diagonal is called the {\em Durfee square}. 
{It} is the maximal square contained in $\lambda$ that includes the cell $(1,1)$. Furthermore, the {\em Durfee rectangle} is defined as the maximal rectangle contained in $\lambda$ that includes $(1,1)$ and whose width exceeds its height by exactly one \footnote{{For a Young diagram with only one column, its Durfee rectangle is defined to have height zero and width one.}} \cite{Andrews1984,Takagi2026}.

A {\em rim hook} is a subset of cells in a Young diagram such that the cells are connected via edges and it contains no $2\times2$ square. Associated with a rim hook are two statistics: the {\em arm length} and the {\em leg length}. The arm length is the number of columns the rim hook occupies minus one, while the leg length is the number of rows it occupies minus one.

We consider mappings $f^{(a)}$ ($a=0,1$) from the set of integer partitions $\mathscr{P}$ to $\mathbb{Z}_{\ge0}$ defined by the procedure described in \cite{Takagi2026}. For a given partition $\lambda \in \mathscr{P}$:

\begin{enumerate}
        \item Let $D_0(\lambda)$ and $D_1(\lambda)$ denote the Durfee square and the Durfee rectangle of $\lambda$, respectively. Let $n_a = n_a(\lambda)$ be the height of $D_a(\lambda)$. We define $A_a(\lambda)$ as the Young diagram obtained by removing $D_a(\lambda)$ from the shape consisting of the first $n_a$ rows of $\lambda$.
        \item From the rim of the diagram $A_a(\lambda)$, remove a rim hook with arm length $n_a+a$, so that the remaining diagram becomes a Young diagram again. Repeat this operation on the remaining diagram until the number of columns in the diagram is less than $n_a+a+1$. Let $R_a(\lambda)$ denote the final diagram.
        \item Let $F(R_a(\lambda))=(x_1,\dots,x_d \mid y_1,\dots,y_d)$. Setting $y_0 = n_a$ and $x_{d+1} = -1$, we define\footnote{{If $R_a(\lambda)=\emptyset$, we set $d=0$ and hence $f^{(a)}(\lambda)=n_a$.}}
        \begin{center}
            $\displaystyle f^{(a)}(\lambda) \coloneq \max_{0\le i\le d}(y_i - x_{i+1}) - 1$.
        \end{center}
\end{enumerate}
%
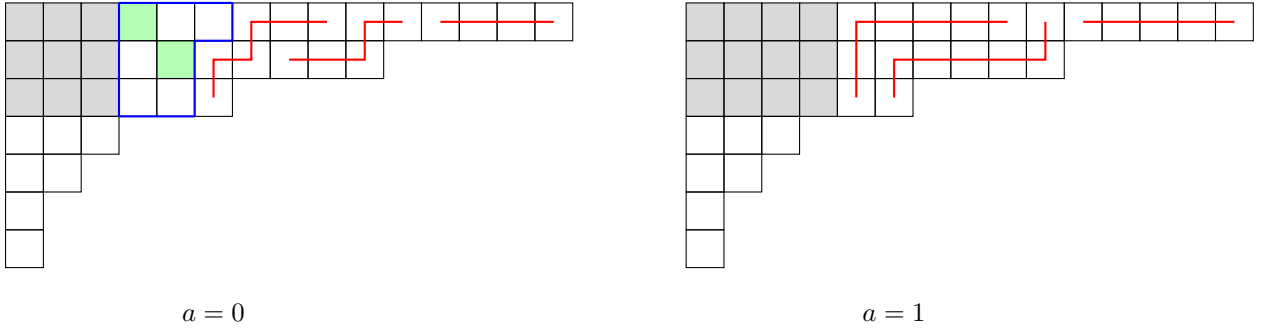
\begin{figure}[htbp] 
  \centering
  \begin{tikzpicture}[x=5mm,y=5mm]
    \def\n{7}
    \colorlet{myblue}{blue}
    
    
    \fill[gray!30] (0,\n) rectangle (3,\n-3);
    \fill[green!30] (3,\n-1) rectangle (4,\n);
    \fill[green!30] (4,\n-2) rectangle (5,\n-1);
    
    \foreach \row/\len in {1/15, 2/10, 3/6, 4/3, 5/2, 6/1, 7/1} {
      \foreach \col in {1,...,\len} {
        \draw (\col-1,\n-\row) rectangle (\col,\n-\row+1);
      }
    }
    
    \draw[red,thick]
      (6-0.5,\n-3+0.5) --
      (6-0.5,\n-2+0.5) --
      (7-0.5,\n-2+0.5) --
      (7-0.5,\n-1+0.5) --
      (8-0.5,\n-1+0.5) --
      (9-0.5,\n-1+0.5);

    \draw[red,thick]
      (8-0.5,\n-2+0.5) --
      (9-0.5,\n-2+0.5) --
      (10-0.5,\n-2+0.5) --
      (10-0.5,\n-1+0.5) --
      (11-0.5,\n-1+0.5);

    \draw[red,thick]
      (12-0.5,\n-1+0.5) --
      (13-0.5,\n-1+0.5) --
      (14-0.5,\n-1+0.5) --
      (15-0.5,\n-1+0.5);
      
    \draw[myblue,thick] (3,\n) -- (6,\n);               
    \draw[myblue,thick] (3,\n) -- (3,\n-3);             
    \draw[myblue,thick] (6,\n) -- (6,\n-1);             
    \draw[myblue,thick] (3,\n-3) -- (5,\n-3);           
    \draw[myblue,thick] (5,\n-1) -- (5,\n-3);           
    \draw[myblue,thick] (5,\n-1) -- (6,\n-1);           
    
    \node[below] at (5.5,-0.7) {$a=0$};

    
    \begin{scope}[xshift=90mm]
      \def\n{7}
      
      \fill[gray!30] (0,\n) rectangle (4,\n-3);
      \foreach \row/\len in {1/15, 2/10, 3/6, 4/3, 5/2, 6/1, 7/1} {
        \foreach \col in {1,...,\len} {
          \draw (\col-1,\n-\row) rectangle (\col,\n-\row+1);
        }
      }

      \draw[red,thick]
        (5-0.5,\n-3+0.5) --
        (5-0.5,\n-2+0.5) --
        (5-0.5,\n-1+0.5) --
        (6-0.5,\n-1+0.5) --
        (7-0.5,\n-1+0.5) --
        (8-0.5,\n-1+0.5) --
        (9-0.5,\n-1+0.5);

      \draw[red,thick]
        (6-0.5,\n-3+0.5) --
        (6-0.5,\n-2+0.5) --
        (7-0.5,\n-2+0.5) --
        (8-0.5,\n-2+0.5) --
        (9-0.5,\n-2+0.5) --
        (10-0.5,\n-2+0.5) --
        (10-0.5,\n-1+0.5);

      \draw[red,thick]
        (11-0.5,\n-1+0.5) --
        (12-0.5,\n-1+0.5) --
        (13-0.5,\n-1+0.5) --
        (14-0.5,\n-1+0.5) --
        (15-0.5,\n-1+0.5);
      
      \node[below] at (5.5,-0.7) {$a=1$};
    \end{scope}
  \end{tikzpicture}
  \caption{The procedure defining the map $f^{(a)}(\lambda)$ for the Young diagram $\lambda=(15,10,6,3,2,1,1)$ with $\text{sqrank}(\lambda)=\max(3-2,2-0,1-(-1))-1=1$ and $\text{rerank}(\lambda)=3$. 
  (Gray area: $D_a(\lambda)$, \blue{area surrounded by blue lines}: $R_0(\lambda)$ ($R_1(\lambda)=\emptyset$), \red{clusters of cells connected by a red line}: rim hooks such that their arm length is equal to $n_a+a$.)
  } \label{fig.lambda}
\end{figure} 

Using the mapping $f^{(a)}$ defined by the above procedure, we define the following statistics for a partition $\lambda$\cite{Takagi2026}:
\begin{equation*}
    \begin{aligned}
        \text{sqrank}(\lambda) &\coloneq f^{(0)}(\lambda), \\
        \text{rerank}(\lambda) &\coloneq f^{(1)}(\lambda).
    \end{aligned}
\end{equation*}

Let $\mathcal{P}_{\Lambda_0}(n, 2r+1)$ be the set of paths in the $U_q(\widehat{\mathfrak{sl}}_2)$-crystal $B(\Lambda_0)$ such that {its arbitrary element} $p$ is an $\tilde{f}_1$ highest {state, with} the energy $E_{\gets}(p)=n$, and $p$ {belongs} to {a} $(2r+1)$-dimensional irreducible representation. 
The cardinality of the set $\mathcal{P}_{\Lambda_0}(n, 2r+1)$ is equal to that of $\mathscr{P}_{\text{sq}}(n,r)$, the set of partitions of $n$ whose {\em sqrank} is equal to $r$.
Similarly, let $\mathcal{P}_{\Lambda_1}(n, 2r'+2)$ be the set of {all the} $\tilde{f}_1$ highest paths in $B(\Lambda_1)$ with {$E_{\gets}(\bullet)=n$} {that belong} to $(2r'+2)$-dimensional irreducible representations.
Then, the cardinality of the set {$\mathcal{P}_{\Lambda_1}(n, 2r'+2)$} is equal to that of $\mathscr{P}_{\text{re}}(n,r')$, the set of partitions of $n$ whose {\em rerank} is equal to $r'$.
The fact that these cardinalities coincide follows from the expression for the branching function (\ref{branchingFunction}) and the explanation below it, Theorem 1 of \cite{Takagi2026}, and the properties of crystal bases \cite{Kashiwara1991}.
We shall give a bijective proof of this fact.

\section{Correspondence Between Young Diagrams \texorpdfstring{$R_a(\lambda)$}{} and Bit Strings} \label{Sect.R}
In what follows, the number of cells in a Young diagram $X$ is called the energy of $X$ and is denoted by $|X|$. Let $\lambda$ be a partition of $n$ such that $\text{sqrank}(\lambda)=r$ and $\text{rerank}(\lambda)=r'$.
The $R_a(\lambda)$ is a Young diagram with at most $n_a$ parts, each of which is at most $n_a+a$; that is, a Young diagram that fits inside the Durfee square/rectangle of $\lambda$.
Let $\mathscr{P}_{n_a+a}^{(n_a)}$ denote the set of all such Young diagrams with at most $n_a$ parts, each of which is at most $n_a+a$.
For $R_a\in\mathscr{P}_{n_a+a}^{(n_a)}$ such that its Frobenius representation is $F(R_a)=(x_1,\dots,x_d\mid y_1,\dots,y_d)$, we define ${\mathcal{F}}(R_a)$ by
\begin{equation*}
    {\mathcal{F}}(\emptyset)=(\underbrace{1,\dots,1}_{n_a}, \underbrace{0, \dots, 0}_{n_a+a}),
\end{equation*}
if $R_a=\emptyset$, and
\begin{equation*}\label{eq:2024oct16_1}
{\mathcal{F}}(R_a)  =(\underbrace{1,\dots,1}_{y_d}, \underbrace{0, \dots, 0}_{x_d+1},
\underbrace{1,\dots,1}_{y_{d-1}-y_d}, \underbrace{0, \dots, 0}_{x_{d-1}-x_d},\dots \dots,\underbrace{1,\dots,1}_{y_{1}-y_2}, \underbrace{0, \dots, 0}_{x_{1}-x_2},
\underbrace{1,\dots,1}_{n_a-y_1}, \underbrace{0, \dots, 0}_{(n_a+a)-x_1-1}),
\end{equation*}
otherwise.
Then, ${\mathcal{F}}(R_a)$ is a bit string {that consists} of $n_a$ ones and $n_a+a$ zeros. Let $\mathcal{H}(2n_a+a,n_a)$ denote the set of all such bit strings.
It is easy to see {that} the map ${\mathcal{F}}$ defined above gives a bijection between $\mathscr{P}_{n_a+a}^{(n_a)}$ and $\mathcal{H}(2n_a+a,n_a)$.
Furthermore, ${\mathcal{F}}$ is an energy-preserving map in the sense that $|R_a|=E_\to({\mathcal{F}}(R_a))$ for any $R_a\in\mathscr{P}_{n_a+a}^{(n_a)}$ and ${\mathcal{F}}(R_a)\in\mathcal{H}(2n_a+a,n_a)${\cite[Sect.~5]{Takagi2026}}.
Now, for $\eta\in\mathcal{H}(2n_a+a,n_a)$, let us consider the path encoding $S(\eta)=(S_i(\eta))_{1\le i\le 2n_a+a}$ defined as follows \cite{CKST2023}:
\begin{align}\label{pathEncoding}
    S_0 (\eta) &= 0,  \nonumber\\
    \quad S_i (\eta) &= S_{i-1} (\eta) +1 - 2 \eta_i \,(1 \leq i \leq 2n_a+a).
\end{align}
We call $S(\eta)$ the path of the bit string $\eta$.
Let $\varepsilon_1(\eta)$ be the absolute value of the minimum of the path $S(\eta)$, given by:
\begin{equation*}
    \varepsilon_1(\eta)\coloneq-\min_{0\le i \le 2n_a+a} S_i(\eta).
\end{equation*}
{This is a standard function in the theory of crystal bases\cite{HK2002, BS2017}.}

For a given Young diagram $\lambda$, consider the bit string $p'_{R_a}(\lambda)\coloneq {\mathcal{F}}(R_a(\lambda))$ corresponding to $R_a(\lambda)$. Then the following holds\cite[Sect.~7]{Takagi2026};
\begin{equation*}
    \begin{aligned}
        \varepsilon_1(p'_{R_0}(\lambda)) &=\max_{0\le i\le d(R_0(\lambda))} (y_i-x_{i+1})-1 =\text{sqrank}(\lambda)=r, \\
        \varepsilon_1(p'_{R_1}(\lambda)) &=\max_{0\le i\le d(R_1(\lambda))} (y_i-x_{i+1})-1=\text{rerank}(\lambda) =r'.
    \end{aligned}
\end{equation*}
Here, $d(R_a(\lambda))$ is the number of diagonal elements of the Young diagram $R_a(\lambda)$, and we set $y_0=n_a$ and $x_{d(R_a(\lambda))+1}=-1$.
Furthermore, it is clear from the definition (\ref{pathEncoding}) that the value of $\varepsilon_1(p'_{R_a}(\lambda))$ remains unchanged even if we repeatedly remove all adjacent 01-pairs in $p'_{R_a}(\lambda)$. 
Therefore, if we let $\tilde{p}'_{R_a}(\lambda)$ denote the bit string obtained by this operation, it takes the form:
\begin{equation*}
    \begin{aligned}
        \tilde{p}'_{R_0}(\lambda)= (\underbrace{1,\dots,1}_{r}, \underbrace{0, \dots, 0}_{r}),\\
        \tilde{p}'_{R_1}(\lambda)= (\underbrace{1,\dots,1}_{r'}, \underbrace{0, \dots, 0}_{r'+1}).
    \end{aligned}
\end{equation*}

Now, let $p_{R_a}(\lambda)$ be the bit string obtained by applying the Kashiwara operator $\tilde{e}_1$ to $p'_{R_a}(\lambda)$ until it becomes a highest weight state with respect to $\tilde{f}_1$. 
Under this operation, the energy is invariant, {which means that} $E_\to(p_{R_a}(\lambda))=E_\to(p'_{R_a}(\lambda))=|R_a(\lambda)|$. 
Repeatedly {removing} all {the }adjacent 01 pairs in $p_{R_0}(\lambda)\,(\text{resp.}\, p_{R_1}(\lambda))$ yields a bit string {that }consists of $2r\,(\text{resp.}\,2r'+1)$ zeros. 
That is, this {belongs} to a $(2r+1)\,(\text{resp.}\,(2r'+2))$-dimensional irreducible representation.

\begin{example}\label{ex.R_a}
As can be seen from Fig.\ref{fig.lambda}, one can obtain the $R_a(\lambda)$ for $\lambda=(15,10,6,3,2,1,1)$ that is the Young diagram shown in Fig.\ref{fig.R_a}. 
Note that $\text{sqrank}(\lambda)=1$ and $\text{rerank}(\lambda)=3$.
%
\begin{figure}[hbtp] 
  \centering
  \begin{tikzpicture}[x=6mm,y=6mm]
    \def\n{7}

    \fill[gray!30] (3,\n) rectangle (6,\n-3);
    \foreach \row in {1,2,3} {
      \foreach \col in {4,5,6} {
        \draw[gray!70,densely dashed] (\col-1,\n-\row) rectangle (\col,\n-\row+1);
      }
    }
    \foreach \row/\start/\len in {1/4/3, 2/4/2, 3/4/2} {
      \foreach \col in {\start,...,\numexpr\start+\len-1\relax} {
        \fill[white] (\col-1,\n-\row) rectangle (\col,\n-\row+1);
      }
    }
    \fill[green!30] (3,\n-1) rectangle (4,\n); 
    \fill[green!30] (4,\n-2) rectangle (5,\n-1); 
    \foreach \row/\start/\len in {1/4/3, 2/4/2, 3/4/2} {
      \foreach \col in {\start,...,\numexpr\start+\len-1\relax} {
        \draw[black] (\col-1,\n-\row) rectangle (\col,\n-\row+1);
      }
    }
    \node[below] at (4.5,3) {$R_0(\lambda)$};

    \begin{scope}[xshift=50mm]
      \fill[gray!30] (4,\n) rectangle (8,\n-3);
      \foreach \row in {1,2,3} {
        \foreach \col in {5,6,7,8} {
          \draw[gray!70,densely dashed] (\col-1,\n-\row) rectangle (\col,\n-\row+1);
        }
      }
      \node[below] at (6,3) {$R_1(\lambda)$};
    \end{scope}
  \end{tikzpicture}
  \caption{$R_a(\lambda)$ for $\lambda=(15,10,6,3,2,1,1).$}
\end{figure}
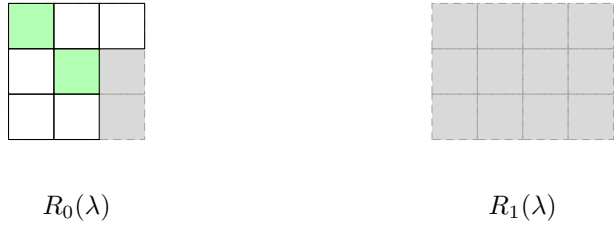 \label{fig.R_a}
%
Their Frobenius representations are $F(R_0(\lambda))=(2,0\mid 2,1)$ and $F(R_1(\lambda))=F(\emptyset)=(\mid)$, respectively. Therefore, we obtain $p'_{R_a}(\lambda)$ in the following way;
\begin{equation*}
    \begin{aligned}
        p'_{R_0}(\lambda)={\mathcal{F}}(R_0(\lambda))&=101001= (\underbracket{\mybox{1}}_{y_2},\overbracket{\mybox{0}}^{x_2+1},\underbracket{\mybox{1}}_{y_1-y_2},\overbracket{\mybox{0,0}}^{x_1-x_2},\underbracket{\mybox{1}}_{n_0-y_1} \overbracket{\mybox{\vphantom{1}}}^{(n_0+0)-x_1-1}),\\
        p'_{R_1}(\lambda)={\mathcal{F}}(R_1(\lambda))&=1110000= (\underbracket{\myboxx{1,1,1}}_{n_1},\overbracket{\myboxxx{0,0,0,0}}^{n_1+1}),
    \end{aligned}
\end{equation*}
and also $\tilde{p}'_{R_a}(\lambda)$, which are obtained by repeatedly canceling all the adjacent 01 pairs therein;
\begin{equation*}
    \begin{aligned}
        \tilde{p}'_{R_0}(\lambda)&=10
        =(\underbracket{\mybox{1}}_{\text{sqrank}(\lambda)},
        \overbracket{\mybox{0}}^{\text{sqrank}(\lambda)}),\\
        \tilde{p}'_{R_1}(\lambda)&=1110000
        =(\underbracket{\myboxx{1,1,1}}_{\text{rerank}(\lambda)},
        \overbracket{\myboxxx{0,0,0,0}}^{\text{rerank}(\lambda)+1}).
    \end{aligned}
\end{equation*}
The bit strings $p_{R_a}(\lambda)$ can be obtained by repeatedly applying $\tilde{e}_1$ to $p'_{R_a}(\lambda)$ until they become highest weight states with respect to $\tilde{f}_1$, and are explicitly given by
\begin{equation*}
    \begin{aligned}
        p_{R_0}(\lambda)&=001001,\\
        p_{R_1}(\lambda)&=0000000,
    \end{aligned}
\end{equation*}
which {belong} to a three-dimensional and {an} eight-dimensional irreducible representation, respectively. Furthermore, we can indeed verify that
\begin{equation*}
    \begin{aligned}
        E_\to(p_{R_0}(\lambda))&= E_\to(p'_{R_0}(\lambda))=|R_0(\lambda)|=7,\\
        E_\to(p_{R_1}(\lambda))&= E_\to(p'_{R_1}(\lambda))=|R_1(\lambda)|=0.
    \end{aligned}
\end{equation*}
    
\end{example}

\section{Correspondence Between Young Diagrams \texorpdfstring{$R_a(\lambda)\uplus D_a(\lambda)$}{} and Bit Strings} \label{Sect.R+D}

In this section, we consider a map from {the set of} bit strings to {themselves} defined by the operation of replacing 01 with 0011.
We define $p_{R_a\uplus D_a}(\lambda)$ as the bit string constructed from $p_{R_a}(\lambda)$ {by this map.}  
{A bit string $p_{R_a\uplus D_a}(\lambda)$ corresponds to a Young diagram $R_a\uplus D_a$ that is obtained by putting $R_a$ just to the right of $D_a$, where $\uplus$ denotes disjoint union of sets.}
For convenience, we write $p_{R_a}(\lambda)\equiv p^{\alpha}\equiv p_{2n_a+a}^{\alpha}\cdots p_2^\alpha p_1^\alpha\in\{0,1\}^{2n_a+a} $, $p_{R_a\uplus D_a}(\lambda)\equiv p^\beta\equiv p_{2n'_a+a}^\beta\cdots p_2^\beta p_1^\beta\in\{0,1\}^{2n'_a+a} $. 
Note that the number of adjacent 01 pairs in $p_{R_a}(\lambda)$ is equal to $n'_a - n_a$. Next, let us consider the energy of the path. By definition (\ref{def:EnergyFunction}) and (\ref{def:EnergyForBitSequence}), 
\begin{equation*}
    \begin{aligned}
        E_\to(p^\alpha)=\sum_{i=1}^{2n_a+a-1}(2n_a+a-i)H_\to(p_{i+1}^\alpha,p_i^\alpha),\\
        E_\to(p^\beta)=\sum_{i=1}^{2n'_a+a-1}(2n'_a+a-i)H_\to(p_{i+1}^\beta,p_i^\beta).
    \end{aligned}
\end{equation*}
Recall that the energy function $H_\to(x,y)$ is not equal to zero if and only if $(x,y)=(0,1)$. 
Therefore, replacing the $k$-th (from the left end) adjacent 01 pair with the sequence 0011 amounts to increasing the right energy by $2k-1$, and we can get the relation;
\begin{equation*}
    \begin{aligned}
        E_\to(p^\beta)=E_\to(p^\alpha)+\sum_{k=1}^{n'_a-n_a}(2k-1)=E_\to(p^\alpha)+(n'_a-n_a)^2.
    \end{aligned}
\end{equation*}
Furthermore, noting that $H_\gets(x,y)=1-H_\to(x,y)$,
\begin{equation*}
    \begin{aligned}
        E_\gets^{\Lambda_a}(p^\beta)&=\sum_{i=1}^{2n'_a+a-1}i[H_\gets(p_{i+1}^\beta,p_i^\beta)-H_\gets(\bar{p}_{i+1}^{\Lambda_a},\bar{p}_i^{\Lambda_a})]\\
        &=\sum_{i=1}^{2n'_a+a-1}i\{[1-H_\to(p_{i+1}^\beta,p_i^\beta)]-[1-H_\to(\bar{p}_{i+1}^{\Lambda_a},\bar{p}_i^{\Lambda_a})]\}\\
        &=\sum_{i=1}^{2n'_a+a-1}(2n'_a+a-i)H_\to(p_{i+1}^\beta,p_i^\beta)-(2n'_a+a)\sum_{i=1}^{2n'_a+a-1}H_\to(p_{i+1}^\beta,p_i^\beta)+\sum_{i=1}^{2n'_a+a-1}iH_\to(\bar{p}_{i+1}^{\Lambda_a},\bar{p}_i^{\Lambda_a})\\
        &=E_\to(p^\beta)-(2n'_a+a)(n'_a-n_a)+n'_a(n'_a+a)\\
        &=E_\to(p^\alpha)+(n'_a-n_a)^2-(2n'_a+a)(n'_a-n_a)+n'_a(n'_a+a)\\
        &=|R_a(\lambda)|+n_a(n_a+a)\\
        &=|R_a(\lambda)|+|D_a(\lambda)|.
    \end{aligned}
\end{equation*}
In addition, it is obvious that {the map that replaces every} adjacent 01 pair with 0011 preserves the dimension of the irreducible {representation to which the bit string belongs}.
It is also clear that this map is {an} {injection}.
{Moreover, this map commutes with the Kashiwara operators $\ft{1}$, $\et{1}$.}

\begin{example}\label{ex.R_a+D_a}
Following Example \ref{ex.R_a}, let us consider $\lambda=(15,10,6,3,2,1,1)$ as an example. Recall that $p_{R_0}(\lambda)=001001$ and $p_{R_1}(\lambda)=0000000$. By replacing each adjacent 2-bit pair 01 with the 4-bit sequence 0011, we obtain
\begin{equation*}
    \begin{aligned}
        p_{R_0\uplus D_0}(\lambda)&=0\underbracket{0011}0\underbracket{0011},\\
        p_{R_1\uplus D_1}(\lambda)&=0{0000}00.
    \end{aligned}
\end{equation*}
These {still belong to} three-dimensional and eight-dimensional irreducible representations, respectively. By calculating the left energy according to Definition (\ref{def:EnergyForBitSequence}), we can verify that
\begin{equation*}
    \begin{aligned}
        E_\gets^{\Lambda_0}(p_{R_0\uplus D_0}(\lambda))&=|R_0(\lambda)|+|D_0(\lambda)|=16,\\
        E_\gets^{\Lambda_1}(p_{R_1\uplus D_1}(\lambda))&=|R_1(\lambda)|+|D_1(\lambda)|=12.
    \end{aligned}
\end{equation*}

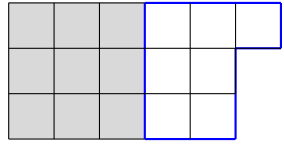
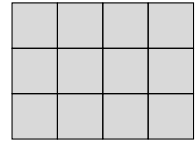
\begin{figure}[hbtp]
  \centering
  \begin{tikzpicture}[x=6mm,y=6mm]
    \def\n{7}
    \colorlet{myblue}{blue}

    \fill[gray!30] (0,\n) rectangle (3,\n-3);
    \draw[myblue,thick] (3,\n) -- (6,\n);
    \draw[myblue,thick] (3,\n) -- (3,\n-3);
    \draw[myblue,thick] (6,\n) -- (6,\n-1);
    \draw[myblue,thick] (3,\n-3) -- (5,\n-3);
    \draw[myblue,thick] (5,\n-1) -- (5,\n-3);
    \draw[myblue,thick] (5,\n-1) -- (6,\n-1);
    \draw (0,\n-3) -- (0,\n);
    \draw (1,\n-3) -- (1,\n);
    \draw (2,\n-3) -- (2,\n);
    \draw (4,\n-3) -- (4,\n);
    \draw (5,\n-2) -- (5,\n);
    \draw (0,\n) -- (3,\n);
    \draw (0,\n-1) -- (6,\n-1);
    \draw (0,\n-2) -- (5,\n-2);
    \draw (0,\n-3) -- (3,\n-3);
    \node[below] at (3,3) {$R_0(\lambda)\uplus D_0(\lambda)$};

    \begin{scope}[xshift=70mm]
      \fill[gray!30] (0,\n) rectangle (4,\n-3);
      \foreach \row in {1,2,3} {
        \foreach \col in {1,2,3,4} {
          \draw (\col-1,\n-\row) rectangle (\col,\n-\row+1);
        }
      }
      \node[below] at (2,3) {$R_1(\lambda)\uplus D_1(\lambda)$};
    \end{scope}
  \end{tikzpicture}
  \caption{$R_a(\lambda)\uplus D_a(\lambda)$ for $\lambda=(15,10,6,3,2,1,1)$.} \label{fig.R_a+D_a}
\end{figure}

\end{example}

\section{Correspondence Between Young Diagrams and Crystal Paths}\label{Sect.lambda}
\subsection{Blocks, Spots, and Strings}\label{Sect.block_spot_string}
We introduce the terms block, spot, and string. Given a bit string, we group 2 bits into pairs starting from the right end.
We call such a pair a {\em block}. 
Specifically, we define a {\em 01-block} as $\mathcal{B}_{01} \coloneq 01$ and a {\em 10-block} as $\mathcal{B}_{10} \coloneq 10$. 
Taking a bit string $p \equiv {p_{2N'+a} \cdots p_2 p_1}$ as an example, a block is given by $B_k \coloneq p_{2k} p_{2k-1}$, and $p$ is written as $p = B_{N'} \cdots B_2 B_1$ or $p = p_{2N'+1} B_{N'} \cdots B_2 B_1$. 
For such a sequence, we define {a} {\em spot} $S_k$ {as a spot just} to the right of block $B_k$. 
Among these spots, those satisfying all of the following ``01-conditions" are called {\em 01-spots}, and those satisfying all of the ``10-conditions" are called {\em 10-spots}.
Note that we have set $p_0=0$.
\begin{flalign*} 
  & \begin{minipage}{0.85\linewidth}
    \begin{itemize}
        \item 01-conditions
            \begin{itemize}
                \item[\textbullet] $B_k\neq \mathcal{B}_{01}$.
                \item[\textbullet] $p_{2k-1}=1$ or $p_{2k-2}=0$.
            \end{itemize}
        \item 10-conditions    
            \begin{itemize}
                \item[\textbullet] $B_k\neq \mathcal{B}_{10}$.
                \item[\textbullet] $p_{2k-1}=0$ or $p_{2k-2}=1$.
            \end{itemize}
    \end{itemize}
  \end{minipage} &&
\end{flalign*}
%
%
{We define {\em string} $\mathfrak{S}_{u,v}$} as a {contiguous array of blocks} $B_u \cdots B_v$ ($1 \le v \le u \le N'$) on the bit string $p=B_{N'} \cdots B_2 B_1$ or $p=p_{2N'+1} B_{N'} \cdots B_2 B_1$ that satisfies all of the following conditions:
\begin{flalign*} 
  & \begin{minipage}{0.85\linewidth}
    \begin{itemize}
      \item For any $k$ satisfying $v\le k\le u$; $B_k \neq \mathcal{B}_{01}$.
      \item For any $k$ satisfying $v< k\le u$; $S_k$ is not 01-spot.
      \item There exists no {contiguous array} $B_{u'} \cdots B_{v'}$ in $p$ such that $B_{u'} \cdots B_{v'} \supsetneq B_u \cdots B_v$ and both of the above two conditions are satisfied.
    \end{itemize}
  \end{minipage} &&
\end{flalign*}
Thus, the string $\mathfrak{S}_{u,v}\coloneq B_u\cdots B_v$ takes one of the following forms, where $m\in\mathbb{Z}_{\ge 0}$.
\begin{flalign} \label{examples:String}
  & \begin{minipage}{0.85\linewidth}
    \begin{itemize}
      \item $00(10)^{m}11$
      \item ${00(10)^{m}}$
      \item ${(10)^{m+1}}$
      \item ${(10)^{m}11}$
    \end{itemize}
  \end{minipage} &&
\end{flalign}
Here, we write ${(10)^{m}} = {10\cdots10} \in \{0,1\}^{2m}$.
It is easy to see that $p_{R_0\uplus D_0}$ and $p_{R_1\uplus D_1}$ {(without its leftmost bit) can} be described using only strings and 01-blocks $\mathcal{B}_{01}$. 
%
%
Moreover, every string in $p_{R_a\uplus D_a}$ is restricted to the form obtained by setting $m=0$ in (\ref{examples:String}).
Now, 
we define the left energy of a string $\mathfrak{S}_{u,v}\equiv p_{2u}p_{2u-1}\cdots p_{2v}p_{2v-1}$ with respect to the ground state {of} $\Lambda_0$ by
\begin{equation*}
    E_{\gets}^{\Lambda_0}(\mathfrak{S}_{u,v})=
    \sum_{j=2v-1}^{2u-1}j[H_{\gets}(p_{j+1},p_j)-H_\gets(\bar{p}_{j+1}^{\Lambda_0},\bar{p}_{j}^{\Lambda_0})].
\end{equation*}
From the definition of the energy of a {finite sequence} (\ref{def:EnergyForBitSequence}), it follows that the left energy of {the} sequence with respect to the ground state $\bar{p}^{\Lambda_0}$ is given by the sum of the left energies of the strings {therein}. This is because, by (\ref{def:GroundState}) and (\ref{def:EnergyFunction}), {one can verify that the} left energy associated with the bits in $\mathcal{B}_{01}$ vanishes {by using the following relations}:
\begin{equation*}
    \begin{aligned}
        H_{\gets}(p_{2k+1},0)-H_{\gets}(\bar{p}^{\Lambda_0}_{2k+1},\bar{p}^{\Lambda_0}_{2k})&=H_{\gets}(p_{2k+1},0)-H_{\gets}(1,0)=0,\\
        H_{\gets}(0,1)-H_{\gets}(\bar{p}^{\Lambda_0}_{2k},\bar{p}^{\Lambda_0}_{2k-1})&=H_{\gets}(0,1)-H_{\gets}(0,1)=0,\\
        H_{\gets}(1,p_{2k-2})-H_{\gets}(\bar{p}^{\Lambda_0}_{2k-1},\bar{p}^{\Lambda_0}_{2k-2})&=H_{\gets}(1,p_{2k-2})-H_{\gets}(1,0)=0.
    \end{aligned}
\end{equation*}
Furthermore, {for all the four types of strings in (\ref{examples:String}) their} left energy with respect to the ground state $\bar{p}^{\Lambda_0}$ is calculated as follows.
\begin{equation} \label{examples:StringEnergy}
    \begin{aligned}
        E_\gets^{\Lambda_0}(\mathfrak{S}_{u,v})=u+v-1\ge 1.
    \end{aligned}
\end{equation}

\subsection{Insertion of Blocks}\label{Sect.Insertion_of_blocks}
Now, let us consider the operation of inserting 01-blocks $\mathcal{B}_{01}$ into 01-spots and 10-blocks $\mathcal{B}_{10}$ into 10-spots of the bit string $p_{R_a\uplus D_a}(\lambda)$ obtained in the previous section. 
In both cases $a=0$ and $a=1$, we first calculate the energy with respect to the ground state of $\Lambda_0$, and for the latter case, we subsequently correct for the energy difference from the ground state of $\Lambda_1$.
Note that the leftmost two bits of this bit string are always {00}.
For $a=1$, there is no difference in {the} energy\footnote{{The same result holds even if the leftmost two bits are 10 or 11.}} between the leftmost 00 part and the 10 in the ground state. Therefore, it suffices to consider only the energy of the remaining bit string excluding the leftmost bit. 
As described in Sect.\ref{Sect.block_spot_string}, both $p_{R_1\uplus D_1}(\lambda)$ with its leftmost bit removed and $p_{R_0\uplus D_0}(\lambda)$ can be described solely in terms of strings and $\mathcal{B}_{01}$, and the bits contained in $\mathcal{B}_{01}$ do not contribute to the energy.
Therefore, 
the energy $E_\gets^{\Lambda_0}(p_{R_a\uplus D_a}(\lambda))$ is given by the sum of the left energies of the strings with respect to the ground state $\bar{p}^{\Lambda_0}$.
From the definitions of strings and 01-spots, it follows that exactly one 01-spot exists immediately to the right of each string, and none exist elsewhere.
On the other hand, exactly one 10-spot exists either left or right side {of} the leftmost block of each string, and none exists elsewhere.
{We refer to such 01- and 10-spots as spots associated with string $\mathfrak{S}_{u,v}$.}
When $\mathcal{B}_{01}$ is inserted into a 01-spot {associated with string $\mathfrak{S}_{u,v}$}, the indices {$u'$ and $v'$} used in the calculation of the energy (\ref{examples:StringEnergy}) for any string {$\mathfrak{S}_{u',v'}$ ($v'\ge v$)} are replaced by {$u'+1$ and $v'+1$}, respectively. {As a result, we see that} {the} left energy of the resulting bit string with respect to the ground state $\bar{p}^{\Lambda_0}$ increases by
\begin{equation*}
    2\times\#\{{\mathfrak{S}_{u',v'}\mid v'\ge v}\}.
\end{equation*}
Next, when $\mathcal{B}_{10}$ is inserted into a 10-spot {associated with string $\mathfrak{S}_{u,v}$}, 
{the indices {$u'$ and $v'$} used in the calculation of the energy (\ref{examples:StringEnergy}) for any string {$\mathfrak{S}_{u',v'}$ ($v'> u$)} are replaced by {$u'+1$ and $v'+1$}, respectively.}
{In addition, the indices {$u$ and $v$} used in the calculation of the energy (\ref{examples:StringEnergy}) for the string {$\mathfrak{S}_{u,v}$} {are} replaced by {$u+1$ and $v$}}
Therefore, the left energy of the bit string obtained by inserting a single $\mathcal{B}_{10}$ increases by
\begin{equation*}
    1+2\times\#\{{\mathfrak{S}_{u',v'}\mid {v' >u}}\}
\end{equation*}
with respect to the state $\bar{p}^{\Lambda_0}$.

{If} $p_{R_a}(\lambda)\in\{0,1\}^{2n_a+a}$ {turns into} $p_{R_a\uplus D_a}(\lambda)\in\{0,1\}^{2n'_a+a}$, {then the latter} bit string $p_{R_a\uplus D_a}(\lambda)$ contains $n'_a-n_a$ adjacent 01 pairs.
From the ``01-conditions", it follows that the number of spots in the bit string that are not 01-spots is equal to the number of adjacent 01 pairs. 
Therefore, the number of 01-spots in $p_{R_a\uplus D_a}(\lambda)$ is $n'_a - (n'_a - n_a) = n_a$. 
In addition, we also find that the number of strings and the number of 10-spots in $p_{R_a\uplus D_a}(\lambda)$ are also equal to $n_a$.
Note that while it is trivial that the dimension of the irreducible representation to which the bit string belongs remains unchanged under the insertion of 01-blocks, the same holds true for 10-blocks. 
This is because the bits surrounding a 10-spot are one of $\{0,0\}$, $\{0,1\}$, or $\{1,1\}$. 
Inserting a 10-block into such a spot yields $\{0,(1,0)^m,0\}$, $\{0,(1,0)^m,1\}$, or $\{1,(1,0)^m,1\}$, respectively. 
However, if we ignore the 01-pairs, these sequences coincide with the original $\{0,0\}$, $\{0,1\}$, and $\{1,1\}$.

From the above discussion, we first insert 01- and 10-blocks into $p_{R_0\uplus D_0}(\lambda)$ such that the energy increases by the number of cells in every removed rim hook and in every row located below the Durfee square in Sect.\ref{subSect.partition}. 
Then, we add the ground state $\bar{p}^{\Lambda_0}$ to the left of the resulting bit string, thereby forming a semi-infinite bit string, namely a path. 
We let $p(\lambda;\Lambda_0)$ denote the uniquely determined path through this procedure, which is an element of $U_q(\widehat{\mathfrak{sl}}_2)$ crystal $B(\Lambda_0)$ corresponding to a given Young diagram $\lambda$ with $\text{sqrank}(\lambda)=r$.
This path is a highest state with respect to $\tilde{f}_1$, satisfies $E_\gets^{\Lambda_0}(p(\lambda;\Lambda_0))=|\lambda|$, and {belongs} to a $(2r+1)$-dimensional irreducible representation.

Next, let us consider the left energy of a path in the crystal $B(\Lambda_1)$ with respect to the ground state $\bar{p}^{\Lambda_1}$.
The left energy of a path $p=p_{2k+1}\cdots p_2p_1$ is given by
\begin{equation*}
    \begin{aligned}
        E_\gets^{\Lambda_1}(p)=& \sum_{j=1}^{2k} j[H_{\gets}(p_{j+1},p_j)-H_{\gets}(\bar{p}_{j+1}^{\Lambda_1},\bar{p}_j^{\Lambda_1})]\\
        =& \sum_{j=1}^{2k} j[H_{\gets}(p_{j+1},p_j)-H_{\gets}(\bar{p}_{j+1}^{\Lambda_0},\bar{p}_j^{\Lambda_0})]
        +\sum_{j=1}^{2k} j[H_{\gets}(\bar{p}_{j+1}^{\Lambda_0},\bar{p}_j^{\Lambda_0})-H_{\gets}(\bar{p}_{j+1}^{\Lambda_1},\bar{p}_j^{\Lambda_1})]\\
        =& E_\gets^{\Lambda_0}(p) + \sum_{j=1}^{2k} j[H_{\gets}(\bar{p}_{j+1}^{\Lambda_0},\bar{p}_j^{\Lambda_0})-H_{\gets}(\bar{p}_{j+1}^{\Lambda_1},\bar{p}_j^{\Lambda_1})]\\
        =& E_\gets^{\Lambda_0}(p) 
        + \sum_{j=1}^{k} (2j-1)[H_{\gets}(\bar{p}_{2j}^{\Lambda_0},\bar{p}_{2j-1}^{\Lambda_0})-H_{\gets}(\bar{p}_{2j}^{\Lambda_1},\bar{p}_{2j-1}^{\Lambda_1})]
        + \sum_{j=1}^{k} (2j)[H_{\gets}(\bar{p}_{2j+1}^{\Lambda_0},\bar{p}_{2j}^{\Lambda_0})-H_{\gets}(\bar{p}_{2j+1}^{\Lambda_1},\bar{p}_{2j}^{\Lambda_1})]\\
        =& E_\gets^{\Lambda_0}(p) + \sum_{j=1}^{k}[-(2j-1)+(2j)] \\
        =& E_\gets^{\Lambda_0}(p) + k.
    \end{aligned}
\end{equation*}
Therefore, each time a $\mathcal{B}_{01}$ or $\mathcal{B}_{10}$ is inserted into $p$, the difference of the energies $k$ in the above equation increases by 1.
Hence, we find that inserting $\mathcal{B}_{01}$ into a 01-spot {associated with string $\mathfrak{S}_{u,v}$} of $p_{R_1\uplus D_1}$ increases the left energy with respect to the ground state $\bar{p}^{\Lambda_1}$ by
\begin{equation*}
    1+2\times\#\{{\mathfrak{S}_{u',v'}\mid v'\ge v\}},
\end{equation*}
while inserting $\mathcal{B}_{10}$ into a 10-spot {associated with string $\mathfrak{S}_{u,v}$} increases it by
\begin{equation*}
    2+2\times\#\{{\mathfrak{S}_{u',v'}\mid v'>u\}}.
\end{equation*}
Furthermore, since $p_{R_1\uplus D_1}$ has the form $00p_{2k-1}\cdots p_2p_1$, inserting $\mathcal{B}_{01}$ immediately to the right of the leftmost bit increases the left energy with respect to $\bar{p}^{\Lambda_1}$ by 1.
\footnote{Therefore, although the position immediately to the right of the leftmost bit of $p_{R_1\uplus D_1}$ is not a spot as defined in Sect.\ref{Sect.block_spot_string}, we will treat it as a spot in what follows.}
\textsuperscript{,}\footnote{{The same result holds even if the leftmost two bits are 10 or 11.}}
This can be verified as follows.
\begin{equation*}
    \begin{aligned}
        E_\gets^{\Lambda_1}(0\bm{01}0&p_{2k-1}\cdots p_2p_1)\\
        =& \sum_{j=1}^{2k+2} j[H_{\gets}(p_{j+1},p_j)-H_{\gets}(\bar{p}_{j+1}^{\Lambda_1},\bar{p}_j^{\Lambda_1})]\\
        =&(2k+2)[H_{\gets}(0,0)-H_{\gets}(0,1)] + (2k+1)[H_{\gets}(0,1)-H_{\gets}(1,0)]\\
        &+ (2k)[H_{\gets}(1,0)-H_{\gets}(0,1)] - (2k)[H_{\gets}(0,0)-H_{\gets}(0,1)]+ E_\gets^{\Lambda_1}(00p_{2k-1}\cdots p_2p_1)\\
        =&1+E_\gets^{\Lambda_1}(00p_{2k-1}\cdots p_2p_1).
    \end{aligned}
\end{equation*}

From the above discussion, we first insert 01- and 10-blocks into $p_{R_1\uplus D_1}(\lambda)$ such that the energy increases by the number of cells in every removed rim hook and in every row located below the Durfee rectangle in Sect.\ref{subSect.partition}. 
Then, we add the ground state $\bar{p}^{\Lambda_1}$ to the left of the resulting bit string, thereby forming a semi-infinite bit string, namely a path. We let $p(\lambda;\Lambda_1)$ denote the uniquely determined path through this procedure, which is an element of $U_q(\widehat{\mathfrak{sl}}_2)$ crystal $B(\Lambda_1)$ corresponding to a given Young diagram $\lambda$ with $\text{rerank}(\lambda)=r'$.
This path is a highest state with respect to $\tilde{f}_1$, satisfies $E_\gets^{\Lambda_1}(p(\lambda;\Lambda_1))=|\lambda|$, and {belongs} to a $(2r'+2)$-dimensional irreducible representation.

\begin{example}\label{ex.lambda}
Similarly to Examples \ref{ex.R_a} and \ref{ex.R_a+D_a}, let us again take $\lambda=(15,10,6,3,2,1,1)$ as an example. We have seen that $p_{R_0\uplus D_0}(\lambda)=00 01 10 00 11$ and $p_{R_1\uplus D_1}(\lambda)=0000000$.
Let $S_{01}^{(e)}$ (resp. $S_{10}^{(e)}$) denote a 01 (resp. 10) spot that increases the left energy with respect to the ground state $\bar{p}^{\Lambda_a}$ by $e$. Then, the spots in these bit strings are as follows:
\begin{equation*}
    \begin{aligned}
        p_{R_0\uplus D_0}(\lambda)&=
        \underbracket{00\,S_{10}^{(1)}}_{string}\,S_{01}^{(2)}01\,
        \underbracket{S_{10}^{(3)}\,10}_{string}\,S_{01}^{(4)}\,
        \underbracket{00\,S_{10}^{(5)}\,11}_{string}\,S_{01}^{(6)},\\
        p_{R_1\uplus D_1}(\lambda)&=
        0\,S_{01}^{(1)}
        \underbracket{00\,S_{10}^{(2)}}_{string}\,S_{01}^{(3)}\,
        \underbracket{00\,S_{10}^{(4)}}_{string}\,S_{01}^{(5)}\,
        \underbracket{00\,S_{10}^{(6)}}_{string}\,S_{01}^{(7)}.
    \end{aligned}
\end{equation*}

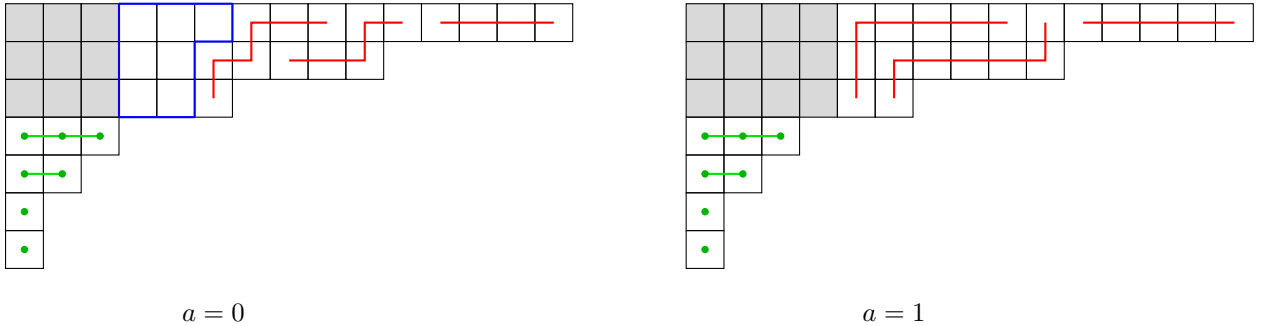
\begin{figure}[htbp]
  \centering
  \begin{tikzpicture}[x=5mm,y=5mm]
    \def\n{7}
    \colorlet{myblue}{blue}

    \fill[gray!30] (0,\n) rectangle (3,\n-3);

    \foreach \row/\len in {1/15, 2/10, 3/6, 4/3, 5/2, 6/1, 7/1} {
      \foreach \col in {1,...,\len} {
        \draw (\col-1,\n-\row) rectangle (\col,\n-\row+1);
      }
    }

    \draw[red,thick]
      (6-0.5,\n-3+0.5) --
      (6-0.5,\n-2+0.5) --
      (7-0.5,\n-2+0.5) --
      (7-0.5,\n-1+0.5) --
      (8-0.5,\n-1+0.5) --
      (9-0.5,\n-1+0.5);

    \draw[red,thick]
      (8-0.5,\n-2+0.5) --
      (9-0.5,\n-2+0.5) --
      (10-0.5,\n-2+0.5) --
      (10-0.5,\n-1+0.5) --
      (11-0.5,\n-1+0.5);

    \draw[red,thick]
      (12-0.5,\n-1+0.5) --
      (13-0.5,\n-1+0.5) --
      (14-0.5,\n-1+0.5) --
      (15-0.5,\n-1+0.5);

    \draw[myblue,thick] (3,\n) -- (6,\n);
    \draw[myblue,thick] (3,\n) -- (3,\n-3);
    \draw[myblue,thick] (6,\n) -- (6,\n-1);
    \draw[myblue,thick] (3,\n-3) -- (5,\n-3);
    \draw[myblue,thick] (5,\n-1) -- (5,\n-3);
    \draw[myblue,thick] (5,\n-1) -- (6,\n-1);
    \draw[green!90!black,thick] (0.5,\n-5+0.5) -- (1.5,\n-5+0.5);
    \draw[green!90!black,thick] (0.5,\n-4+0.5) -- (2.5,\n-4+0.5);
    \fill[green!70!black] (0.5,\n-4+0.5) circle (1.4pt);
    \fill[green!70!black] (1.5,\n-4+0.5) circle (1.4pt);
    \fill[green!70!black] (2.5,\n-4+0.5) circle (1.4pt);
    \fill[green!70!black] (0.5,\n-5+0.5) circle (1.4pt);
    \fill[green!70!black] (1.5,\n-5+0.5) circle (1.4pt);
    \fill[green!70!black] (0.5,\n-6+0.5) circle (1.4pt);
    \fill[green!70!black] (0.5,\n-7+0.5) circle (1.4pt);

    \node[below] at (5.5,-0.7) {$a=0$};

    \begin{scope}[xshift=90mm]
      \def\n{7}
      \fill[gray!30] (0,\n) rectangle (4,\n-3);

      \foreach \row/\len in {1/15, 2/10, 3/6, 4/3, 5/2, 6/1, 7/1} {
        \foreach \col in {1,...,\len} {
          \draw (\col-1,\n-\row) rectangle (\col,\n-\row+1);
        }
      }

      \draw[red,thick]
        (5-0.5,\n-3+0.5) --
        (5-0.5,\n-2+0.5) --
        (5-0.5,\n-1+0.5) --
        (6-0.5,\n-1+0.5) --
        (7-0.5,\n-1+0.5) --
        (8-0.5,\n-1+0.5) --
        (9-0.5,\n-1+0.5);

      \draw[red,thick]
        (6-0.5,\n-3+0.5) --
        (6-0.5,\n-2+0.5) --
        (7-0.5,\n-2+0.5) --
        (8-0.5,\n-2+0.5) --
        (9-0.5,\n-2+0.5) --
        (10-0.5,\n-2+0.5) --
        (10-0.5,\n-1+0.5);

      \draw[red,thick]
        (11-0.5,\n-1+0.5) --
        (12-0.5,\n-1+0.5) --
        (13-0.5,\n-1+0.5) --
        (14-0.5,\n-1+0.5) --
        (15-0.5,\n-1+0.5);

      \draw[green!90!black,thick] (0.5,\n-5+0.5) -- (1.5,\n-5+0.5);
      \draw[green!90!black,thick] (0.5,\n-4+0.5) -- (2.5,\n-4+0.5);

      \fill[green!70!black] (0.5,\n-4+0.5) circle (1.4pt);
      \fill[green!70!black] (1.5,\n-4+0.5) circle (1.4pt);
      \fill[green!70!black] (2.5,\n-4+0.5) circle (1.4pt);
      \fill[green!70!black] (0.5,\n-5+0.5) circle (1.4pt);
      \fill[green!70!black] (1.5,\n-5+0.5) circle (1.4pt);
      \fill[green!70!black] (0.5,\n-6+0.5) circle (1.4pt);
      \fill[green!70!black] (0.5,\n-7+0.5) circle (1.4pt);
      \node[below] at (5.5,-0.7) {$a=1$};
    \end{scope}
  \end{tikzpicture}
  \caption{Young diagram $\lambda=(15,10,6,3,2,1,1)$. 
  (A cluster of $e$ cells connected by a \green{green} (resp. \red{red}) line corresponds to a block increasing the left energy by $e\le n_a(\lambda)+a$ (resp. $e> n_a(\lambda)+a$).)
  } \label{fig.lambda2}
\end{figure}
By inserting 01 and 10 blocks according to Fig.\ref{fig.lambda2}, we obtain
\begin{equation*}
    \begin{aligned}
        p(\lambda;\Lambda_0)&=
        \underbracket{00\green{1010}}_{string}
        \green{01}01
        \underbracket{\green{10}10}
        \red{01}
        \underbracket{00\red{10}11}\red{01},\\
        p(\lambda;\Lambda_1)&=
        0\green{0101}
        \underbracket{00\green{10}}_{string}\green{01}
        \underbracket{00}\red{01}
        \underbracket{00}\red{0101},
    \end{aligned}
\end{equation*}
and we can indeed verify that
\begin{equation*}
    \begin{aligned}
        E_{\gets}^{\Lambda_0}(p(\lambda;\Lambda_0))&=|\lambda|=38,\\
        E_{\gets}^{\Lambda_1}(p(\lambda;\Lambda_1))&=|\lambda|=38,
    \end{aligned}
\end{equation*}
holds. Furthermore, these still {belong} to a three-dimensional and {an} eight-dimensional irreducible representation, respectively.
    
\end{example}

\section{Recovering Young Diagrams from Paths}\label{Sect.Reverse}
We wish to show that the above map from partitions to paths is a bijection. Due to the coincidence of the cardinalities mentioned at the end of Sect.\ref{Sect.2}, it suffices to show that the map is injective. In what follows, we shall demonstrate this.
Given a path $p(\lambda;\Lambda_a)$ in the crystal $B(\Lambda_a)$ of $U_q(\widehat{\mathfrak{sl}}_2)$ that is a highest state with respect to $\tilde{f}_1$, has left energy $E_\gets^{\Lambda_a}(p(\lambda;\Lambda_a))=n$, and {belongs} to a $(2r+1+a)$-dimensional irreducible representation, let us consider how to obtain the corresponding Young diagram $\lambda$.
First of all, from a given path, we can determine whether $a=0$ or $1$ by observing its behavior sufficiently far to the left, and we can also determine the vertical edge length of $D_a$ from the number of strings it contains.

Let us verify that $p_{R_a\uplus D_a}(\lambda)$ can be uniquely obtained from the path $p(\lambda;\Lambda_a)$.
For a block $B_{k}=p_{2k}p_{2k-1}$ in a bit string $p=p_N\cdots p_2p_1$, we consider the following conditions; 
\begin{flalign*} \label{blockRemovingConditions}
  & \begin{minipage}{0.85\linewidth}
    \begin{itemize}
        \item 01-condition
            \begin{itemize}
                \item[\circnum{1}] $B_{k+1}\neq \mathcal{B}_{01}$.
                \item[\circnum{2}] $p_{2k+1}=1$ or $p_{2k-2}=0$.
            \end{itemize}
        \item 10-{condition}   
            \begin{itemize}
                \item[\circnum{1}] $B_{k+1}\neq \mathcal{B}_{10}$.
                \item[\circnum{2}] $p_{2k+1}=0$ or $p_{2k-2}=1$.
            \end{itemize}
    \end{itemize}
  \end{minipage} &&
\end{flalign*}
When a block $\mathcal{B}_{01}$ (resp. $\mathcal{B}_{10}$) in a bit string satisfies the 01-condition (resp. 10-condition) \circnum{1}, we say that the block is {\em determinate}, and we can decide whether or not it can be removed from the bit string. On the other hand, when it does not satisfy this condition, we say that the block is {\em indeterminate}, and we cannot decide whether it can be removed.
When a block $\mathcal{B}_{01}$ (resp. $\mathcal{B}_{10}$) is determinate and furthermore satisfies the 01-condition (resp. 10-condition) \circnum{2}, we say that the block is {\em removable}, and it can be removed from the bit string. On the other hand, when it is determinate but does not satisfy this condition, we say that the block is {\em irremovable}, and it cannot be removed from the bit string.

Now, suppose we successively remove all removable blocks $\mathcal{B}_{01}$ and $\mathcal{B}_{10}$ from an $\tilde{f}_1$-highest path $p(\lambda;\Lambda_a)\in B(\Lambda_a)$ {until there remain} neither removable nor indeterminate blocks.
Then, every occurrence of adjacent bits `01' in the resulting {bit string} must be in the form of `0011'.
This {comes immediately} from a case-by-case analysis. 
{Suppose that the} `01' constitutes a single block.
{Then since} it must be an irremovable block, {it inevitably} takes the form `0011'.
{Suppose that the} `01' spans two blocks.
{Then the blocks look like 0010, 1011, 1010, or 0011. However,} the right block of `0010' and the left block of `1011' {are removable $\mathcal{B}_{10}$s, and} the right block of `1010' is {an indeterminate $\mathcal{B}_{10}$}. 
Therefore, the only permitted configuration is `0011'.
{Conversely, if every occurrence of adjacent bits `01' in a bit string is in the form of `0011', then there exist neither removable nor indeterminate blocks in the bit string. This is because if a bit string has such blocks, then the bit string must hold either `010' or `101' in it.}
%
%
Thus, through this operation, we can uniquely recover $p_{R_a\uplus D_a}(\lambda)$ from $p(\lambda;\Lambda_a)$.
Then, by reversing the operations in Sect.\ref{Sect.R}-\ref{Sect.R+D}, we obtain $R_a(\lambda)$, which immediately yields {$R_a(\lambda)\uplus D_a(\lambda)$.
}
By keeping track of the positions and the number of blocks removed during the process of obtaining $p_{R_a\uplus D_a}(\lambda)${, which may be encoded in a partition with each part $\leq 2n_a+a$,} we can recover the Young diagram $\lambda$ corresponding to $p(\lambda;\Lambda_a)$ based on this {partition} by using the bijection constructed in Proposition 12 and Corollary 13 of \cite{Takagi2026}.

\begin{example}
Here, let us look at the process of obtaining $p_{R_a\uplus D_a}(\lambda)$ from $p(\lambda;\Lambda_0)=001010 0101 1010 01 001011 01$ and $p(\lambda;\Lambda_1)=0 0101 0010 01 00 01 00 0101$. This process is illustrated in Fig.\ref{fig.removingBlocks}.
\begin{figure}[htbp]
\centering
\begin{equation*}
    \begin{aligned}
        p(\lambda;\Lambda_0)&=
        00 \green{\underbracket{\black{10}}} \underbracket{10} 
        \green{\underbracket{\black{01}}} \underbracket{01}
        \green{\underbracket{\black{10}}} \underbracket{10}
        \green{\underbracket{\black{01}}}
        00 \green{\underbracket{\black{10}}} 11
        \green{\underbracket{\black{01}}}\\
        &\to 00 \green{\underbracket{\black{10}}}\, 
        \red{\underbracket{\black{01}}}\,
        \red{\underbracket{\black{10}}} 00 11\\
        &\to 00 \red{\underbracket{\black{01}}}\,
        \red{\underbracket{\black{10}}} 00 11 =p_{R_0\uplus D_0}(\lambda)\\\\
        p(\lambda;\Lambda_1)&=0 
        \green{\underbracket{\black{01}}} \underbracket{01} 
        00\green{\underbracket{\black{10}}} 
        \green{\underbracket{\black{01}}} 
        00 \green{\underbracket{\black{01}}}
        00
        \green{\underbracket{\black{01}}} \underbracket{01}\\
        &\to 0 \green{\underbracket{\black{01}}} 00 00 00 \green{\underbracket{\black{01}}}\\
        &\to 0 00 00 00 =p_{R_1\uplus D_1}(\lambda)
    \end{aligned}
\end{equation*}   
$\green{\underbracket{\black{\bullet\bullet}}}$ : removable block,\quad
$\red{\underbracket{\black{\bullet\bullet}}}$ : irremovable block,\quad
$\underbracket{\black{\bullet\bullet}}$ : indeterminate block.\medskip
\caption{Removing blocks.} \label{fig.removingBlocks}
\end{figure}

\end{example}

%
%

\section{Summary and Discussions}\label{Sect.Summary_Discussions}
\subsection{A summary of the present work}
In this paper, we have developed an explicit procedure to {define} novel energy preserving maps
$p(\bullet; \Lambda_a) : \mathscr{P}  \rightarrow  B(\Lambda_a)$ for $a=0,1$
by using manipulations on Young diagrams and the other techniques in combinatorics.
Let us summarize the arguments we have used there.

For every partition $\lambda \in \mathscr{P}$ there is a unique decomposition \cite[Sect.~7]{Takagi2026}
\begin{equation}\label{eq:2024dec6_1}
\lambda = D_a(\lambda) \uplus A_a(\lambda) \uplus L_a(\lambda).
\end{equation}
Here $A_a(\lambda) \in \mathscr{P}^{({n_a})}$ and
$L_a(\lambda) \in \mathscr{P}_{{n_a}+a}$ are the ``wing" and the ``leg" in Sect.~\ref{sec1}
with ${n_a} = n_a(\lambda)$.
In addition, there is an energy preserving bijection
$\Psi^{(a)} : \mathscr{P}^{({n_a})} \rightarrow \mathscr{P}^{({n_a})}_{{n_a}+a} \times (\mathscr{P}_{2{n_a}+a} \setminus \mathscr{P}_{{n_a}+a})$ \cite[Proposition 12]{Takagi2026}.
We let $\Psi^{(a)}_1$ and $\Psi^{(a)}_2$ be the pair of maps 
obtained from $\Psi^{(a)}$ by composing with a projection
into the first and the second component of $\mathscr{P}^{({n_a})}_{{n_a}+a} \times 
(\mathscr{P}_{2{n_a}+a} \setminus \mathscr{P}_{{n_a}+a})$,
respectively.
Then the restricted partition $R_a(\lambda)$ 
defined in Sect.~\ref{subSect.partition} is given by 
$R_a(\lambda) = \Psi^{(a)}_1(A_a(\lambda)) \in \mathscr{P}^{({n_a})}_{{n_a}+a}$.
Let us consider another partition
\begin{equation}\label{eq:2026may13_1}
Q_a(\lambda) :=\Psi_2^{(a)} (A_a(\lambda)) \cup  L_a(\lambda) \in \mathscr{P}_{2{n_a}+a},
\end{equation}
which encodes the data on
the number of cells in every removed rim hook and in every row located below the Durfee square/rectangle in Sect.~\ref{Sect.R}. 

For every partition $\lambda$, we defined
$p'_{R_a}(\lambda) = {\mathcal{F}}(R_a(\lambda))$ in Sect.~\ref{Sect.R}, which is
a bit string that consists of $n_a$ ones and $n_a+a$ zeros. 
Then we defined 
\begin{equation}\label{eq:2026apr27_1}
p_{R_a}(\lambda) = \et{1}^{f^{(a)} (\lambda)} p'_{R_a}(\lambda),
\end{equation}
which is a $\ft{1}$-highest element.
Here $\et{1}, \ft{1}$ are the Kashiwara operators, 
$f^{(a)} (\lambda)$ is the sqrank/rerank of $\lambda$ defined in Sect.~\ref{subSect.partition}, 
and we regard bit strings
$p'_{R_a}(\lambda)$ and $p_{R_a}(\lambda)$ as elements of 
the tensor product of crystals $B^{\otimes 2 n_a + a}$.
In Sect.~\ref{Sect.R+D},
by applying the operation that replaces every adjacent $01$ pair with $0011$ on $p_{R_a}(\lambda)$,
we obtained a longer bit string $p_{R_a \uplus D_a}(\lambda)$ that is also a $\ft{1}$-highest element.
The notions of blocks, spots, and strings were introduced in Sect.~\ref{Sect.block_spot_string}.
The {\em strings} are particular {contiguous arrays} of blocks 
{lying} in the path realization of the elements of $B(\Lambda_a)$ 
that are classified into the four types shown in \eqref{examples:String}.
The bit {sequence} $p_{R_a \uplus D_a}(\lambda)$ viewed as an element of $B(\Lambda_a)$ has
$n_a$ {such} strings.
Each string has one $10$-spot and one $01$-spot.
In addition, for the case of $a=1$, 
the bit string $p_{R_1 \uplus D_1}(\lambda)$ has
an extra $01$-spot to the left of the leftmost string in $p_{R_1 \uplus D_1}(\lambda)$.
So, in general a bit string $p_{R_a \uplus D_a}(\lambda)$ has $2n_a + a$ such spots in total.
In {Sect.~\ref{Sect.Insertion_of_blocks}} we defined a procedure of insertions of
$10$/$01$-blocks into these $2n_a + a$ spots
by using the data encoded in the partition $Q_a(\lambda)$ {given by \eqref{eq:2026may13_1}},
which may be represented as $1^{m_1} 2^{m_2} \dots (2{n_a}+a)^{m_{2{n_a}+a}}$
where $m_k$ denotes the multiplicity of part $k \in \{1, \dots, 2{n_a}+a\}$ {therein lies}.
In the case of $a=0$, we insert $m_{2k-1}$ $10$-blocks into the $k$-th $10$-spot,
and insert $m_{2k}$ $01$-blocks into the $k$-th $01$-spot of $p_{R_0 \uplus D_0}(\lambda)$,
where the order of spots are defined along the bit string
from the left to the right.
On the other hand,
in the case of $a=1$, we insert $m_{2k-1}$ $01$-blocks into the $k$-th $01$-spot,
and insert $m_{2k}$ $10$-blocks into the $k$-th $10$-spot of $p_{R_1 \uplus D_1}(\lambda)$.
After this procedure,  we obtain a $\ft{1}$-highest path $p(\lambda; \Lambda_a) \in B(\Lambda_a)$
with the property $E_\gets^{\Lambda_a}(p(\lambda;\Lambda_a))=|\lambda|$ where
the definition of the energy $E_\gets^{\Lambda_a}$ was given in \eqref{def:EnergyForPath}.

As we discussed in Sect.~\ref{Sect.Reverse},
every procedure introduced so far to obtain $p(\lambda; \Lambda_a)$ from $\lambda$ is invertible, so
the maps that send partitions $\lambda$ to paths $p(\lambda;\Lambda_a)$ actually
provide bijections between the set of all partitions $\mathscr{P}$ and
the set of all $\ft{1}$-highest elements of the affine crystal $B(\Lambda_a)$ for $a =0,1$.
More precisely, 
in terms of their particular subsets introduced in Sect.~\ref{subSect.partition}, one observes that
the maps we obtained turn out to be 
a bijection between $\mathscr{P}_{\text{sq}}(n,r)$ and $\mathcal{P}_{\Lambda_0}(n, 2r+1)$
for $a=0$, and 
a bijection between $\mathscr{P}_{\text{re}}(n,r')$ and $\mathcal{P}_{\Lambda_1}(n, 2r'+2)$
for $a=1$.

\subsection{Another viewpoint for the bijections}
{As we have already noted, by ignoring} all the arrows for $\ft{0}$, the crystal graphs $B(\Lambda_a)$ {decompose} into
{an} infinite number of connected components of graphs that consist of {a} finite number of
the nodes for paths and the arrows for $\ft{1}$.
They are crystal graphs for the crystal bases of the
irreducible finite dimensional $U_q(\mathfrak{sl}_2)$ modules.
The bijections we {have constructed so far} may be viewed as bijections between 
the set of all partitions $\mathscr{P}$ and the set of all such connected components of crystal graphs,
in the following sense.
Let $\mathcal{C}_{\Lambda_a}(n, 2r+1+a)$ be the set of all such connected components
in $B(\Lambda_a)$ with exactly $2r+1+a$ nodes and whose each node corresponds to a path $p$
with the energy $E_\gets^{\Lambda_a}(p)=n$.
Regard each connected component as an equivalence class in which its nodes belong to.
Then the maps we obtained can be viewed as 
a bijection between $\mathscr{P}_{\text{sq}}(n,r)$ and $\mathcal{C}_{\Lambda_0}(n, 2r+1)$
for $a=0$, and as
a bijection between $\mathscr{P}_{\text{re}}(n,r')$ and $\mathcal{C}_{\Lambda_1}(n, 2r'+2)$
for $a=1$.
Here, the $\ft{1}$-highest elements are regarded as their representatives.
We {shall present} another choice for the representatives.

{From the bit sequence $p'_{R_a}(\lambda)$ defined in Sect.~\ref{Sect.R}, we can obtain
another bit sequence}
by applying the operation that replaces every adjacent $01$ pair with $0011$.
The {resulting bit sequence, which would be denoted by $p'_{R_a \uplus D_a}(\lambda)$, 
still holds}
$n_a$ strings {within it}.
If we apply Kashiwara operator $\et{1}$ on $p'_{R_a \uplus D_a}(\lambda)$,
it can only change a bit in one of the strings.
{More precisely},  $\et{1}$ changes either a string of type $(10)^{m+1}$ into that of type $00 (10)^m$,
or a string of type $(10)^m 11$ into that of type $(10)^{m+1}$, unless it deletes the
element.
It is easy to see that the operation of replacing every adjacent $01$ pair with $0011$
and the action of Kashiwara operator commutes, so from \eqref{eq:2026apr27_1} one observes
the property
\begin{equation}\label{eq:2026apr27_2}
p_{R_a \uplus D_a}(\lambda) = \et{1}^{f^{(a)} (\lambda)} p'_{R_a \uplus D_a}(\lambda).
\end{equation}
As in the case of $p_{R_a \uplus D_a}(\lambda)$, the number of 10-spots in
the bit string $p'_{R_a \uplus D_a}(\lambda)$ is $n_a$, and that of 01-spots is $n_a + a$.
One can apply the same procedure of insertions of $10$/$01$-blocks
into these $2n_a + a$ spots in $p'_{R_a \uplus D_a}(\lambda)$, 
by using the data encoded in the partition $Q_a(\lambda) \in \mathscr{P}_{2{n_a}+a}$.
After this procedure,  we obtain a path $p'(\lambda; \Lambda_a) \in B(\Lambda_a)$
with the properties $E_\gets^{\Lambda_a}(p'(\lambda;\Lambda_a))=|\lambda|$ and
$\varphi_1(p'(\lambda;\Lambda_a)) - \varepsilon_1(p'(\lambda;\Lambda_a)) = a$.
Here $\varphi_1, \varepsilon_1$ are functions in the theory of crystals that are
sometimes referred to as string lengths \cite{HK2002, BS2017}.
It is also easy to see that the procedure of insertions of $10$/$01$-blocks
and the action of Kashiwara operator commutes, so from \eqref{eq:2026apr27_2} one observes
the property
\begin{equation*}\label{eq:2026apr27_3}
p(\lambda;\Lambda_a) = \et{1}^{f^{(a)} (\lambda)} p'(\lambda;\Lambda_a).
\end{equation*}
Therefore $p(\lambda;\Lambda_a)$ and $p'(\lambda;\Lambda_a)$
belong to the same connected component and that enables us to adopt
the latter as another representative.

\subsection{Relation to the spinon representation}
For a potential application of the present work to problems in mathematical physics,
we want to discuss relations between the spinon representations of $\widehat{\mathfrak{sl}}_2$
by Bernard et.~al.~\cite[Sect.~3]{BPS1994} and the notion of strings in Sect.~\ref{Sect.block_spot_string}. 
As a subset of $B(\Lambda_a)$ we define
\begin{equation*}
\mathcal{P}_{\Lambda_a} :=\bigsqcup_{n \geq 0} \{ p \in B(\Lambda_a) \mid \mbox{\textit{wt}} (p) = \Lambda_a - n \delta \},
\end{equation*}
where $\mbox{\textit{wt}} (p)$ denotes the affine weight of $p$.
We note that the procedure to construct
the paths $p'(\lambda;\Lambda_a)$ in the previous subsection
provides energy preserving bijections 
$p'(\bullet;\Lambda_a): \mathscr{P} \rightarrow \mathcal{P}_{\Lambda_a}$
for $a=0,1$.
The spinon representation gives alternative energy preserving bijections between these sets.

{We propose a precise interpretation of the motif description of spinons suggested
in \cite[Sect.~3]{BPS1994} as follows. 
Recall that there are four types of strings in Sect.~\ref{Sect.block_spot_string}.
A spinon can exist between adjacent bits in our bit sequences, and there are two spinons lying within each type of the strings.}
%
Given a string,
one spinon exists between the leftmost bit and its right neighboring bit, and 
the other spinon exists to the right of the rightmost bit, of the string.
If its left neighboring bit is $0$ (resp. $1$), then the spinon is an up spinon $\green{\uparrow}$ (resp. a down spinon $\green{\downarrow}$).
Therefore, for generic cases the spinons associated with the strings can be
illustrated in the following way,
\begin{equation}\label{examples:Stringxxxxxx}
0 \green{\uparrow} 0(10)^{m}11 \green{\downarrow},\qquad
0 \green{\uparrow} 0(10)^{m} \green{\uparrow},\qquad
1 \green{\downarrow}0 (10)^{m} \green{\uparrow},\qquad
1 \green{\downarrow}0 (10)^{m-1}11 \green{\downarrow},
\end{equation}
and for the strings with only one block, their spinons are shown as
$0 \green{\uparrow} 0 \green{\uparrow} , 1\green{\downarrow}0 \green{\uparrow}$ and
$1\green{\downarrow}1\green{\downarrow}$.

For every
$p = \cdots p_2 p_1 \in \mathcal{P}_{\Lambda_a}$, one can specify its string content.
Suppose $p$ has $N$ strings $\{ \mathfrak{S}_{u_i,v_i} \}_{1 \leq i \leq N}$,
where we numbered them increasingly from the right to the left.
This implies that the condition $u_N \geq v_N >u_{N-1} \geq v_{N-1} > \cdots > u_1 \geq v_1 >0$
is satisfied.
There are $2N$ spinons in $p$.
For the $i$-th string $\mathfrak{S}_{u_i,v_i}$, we associate its left spinon with momentum $k_{2i} :=u_i-i$ and
its right spinon with momentum $k_{2i-1}:=v_i-i$.
Then we have $k_{2N} \geq k_{2N-1} \geq \dots \geq k_1 \geq 0$.

We begin with the case of $a=0$.
By using \eqref{examples:StringEnergy}
the energy of $p \in \mathcal{P}_{\Lambda_0}$ is given by
\begin{equation}\label{eq:2026apr28_1}
E_\gets^{\Lambda_0}(p)= \sum_{i=1}^N (u_i + v_i -1) = \sum_{i=1}^{N} (u_i-i + v_i -i)+ \sum_{i=1}^N (2i-1)
= \sum_{i=1}^{2N} k_i + N^2.
\end{equation}
Next we consider the case of $a=1$.
Then the path takes the form $p = \cdots 0101 p_{2m+1} p_{2m} \cdots p_1 \in \mathcal{P}_{\Lambda_1}$
where $m \geq 0$ is the smallest integer satisfying
$(p_{2m+1}, p_{2m}) \ne (0,1)$.
As was shown in Sect.~\ref{Sect.Insertion_of_blocks} one sees that
$E_\gets^{\Lambda_1}(p) = E_\gets^{\Lambda_0}(p) + m$.
Let $v_{N+1} = m+1$ and regard the semi-infinite sequence $\cdots 0101 p_{2m+1}$ as a
``string without left end", which may be denoted by $\mathfrak{S}_{v_{N+1}}$.
To the right of its rightmost bit $p_{2m+1}$, we associate its right 
spinon with momentum $k_{2N+1} :=v_{N+1} -(N+1)$.
Then we have $k_{2N+1} \geq k_{2N}$ and the energy of $p \in \mathcal{P}_{\Lambda_1}$ is given by
\begin{equation}\label{eq:2026apr28_2}
E_\gets^{\Lambda_1}(p)= m+ \sum_{i=1}^N (u_i + v_i -1) = 
\sum_{i=1}^{N} (u_i-i) +\sum_{i=1}^{N+1}  (v_i -i)+N+ \sum_{i=1}^N (2i-1)
= \sum_{i=1}^{2N+1} k_i + N(N+1).
\end{equation}
The expressions \eqref{eq:2026apr28_1} and \eqref{eq:2026apr28_2} coincide up to a constant with the spinon description
of the Virasoro generator $\mathsf{L}_0$ for Wess-Zumino-Witten conformal field theory models
proposed by Bernard et.~al.~\cite[Sect.~3]{BPS1994},
corresponding to the vacuum and the spin half representations respectively.

The alternative energy preserving bijections between $\mathscr{P}$ and $\mathcal{P}_{\Lambda_a}$
are described as follows.
Once again, we begin with the case of $a=0$.
Since every $p \in \mathcal{P}_{\Lambda_0}$ satisfies the relation 
$\varphi_1(p) - \varepsilon_1(p) = 0$, the number of up spinons and that of down spinons
in $p$ must coincide with each other.
{This is because the number of the second and the fourth types in
\eqref{examples:Stringxxxxxx} of the strings in $p$ must coincide.}
Suppose $p$ has up spinons 
with momenta $\{ \lambda^+_1 \geq \lambda^+_2 \geq \dots \geq \lambda^+_N \geq 0\}$, and down spinons
with momenta $\{ \lambda^-_1 \geq \lambda^-_2 \geq \dots \geq \lambda^-_N \geq 0\}$.
We let $\lambda^+ = (\lambda^+_1, \dots , \lambda^+_N)$ and $\lambda^- = (\lambda^-_1, \dots , \lambda^-_N)$ be a pair of partitions with at most $N$ parts.
Hence by \eqref{eq:2026apr28_1} we have $E_\gets^{\Lambda_0}(p)=|\lambda^+| + |\lambda^-| + N^2$.
According to the decomposition \eqref{eq:2024dec6_1} for a partition, say $\pi$,
we let $D_0(\pi)$ be an $N \times N$ square and let 
$A_0(\pi)=\lambda^-, L_0(\pi) = (\lambda^+)'$. Then
we have $\pi = D_0(\pi) \uplus A_0(\pi) \uplus L_0(\pi)$ that has the property
$E_\gets^{\Lambda_0}(p)=|\pi|$.
{This procedure for constructing} $\pi$ from $p$ is invertible.
It is easy to see that one can retrieve the pair of partitions $\lambda^+$ and
$\lambda^-$ from $\pi$.
The spinon momentum content is given by
$\{ k_{2N}, \dots, k_1 \} = \lambda^+ \cup \lambda^-$ where
$k_i \geq k_j$ for every $i > j$.
Let $\sigma_i$ be the sign of $i$-th spinon 
{with momentum $k_i$}, which is determined as follows.
For the above partitions $\lambda^+, \lambda^-$ and every nonnegative integer $k$,
we define $a_k = \#\{ i \mid \lambda^+_i = k \}$ and $b_k = \#\{ i \mid \lambda^-_i = k \}$.
If $a_k \ne 0, b_k =0$ (resp.~$a_k = 0, b_k \ne 0$) we set $\sigma_i = +$ (resp.~$\sigma_i = -$) for every $i$ satisfying $k_i = k$.
Suppose $a_k b_k \ne 0$ and let $i \geq 0$ be the integer
that is uniquely determined by $k_{i+1}=k_{i+2}=\dots=k_{i +a_k+b_k} = k$.
We set $\sigma_j = +$ (resp.~$\sigma_j = -$) for $i < j \leq i+a_k$ (resp.~$i+a_k < j \leq i+a_k+b_k$).
The positions of the left and right ends of $i$-th string are given by 
$u_i = k_{2i}+i, v_i=k_{2i-1}+i$ for $1 \leq i \leq N$.
First we suppose $u_i > v_i$ for some $i$.
If $\sigma_{2i} = +$ (resp.~$\sigma_{2i} = -$) then we set $B_{u_i} = 00$ (resp.~$B_{u_i} = 10$), and 
if $\sigma_{2i-1} = +$ (resp.~$\sigma_{2i-1} = -$) then we set $B_{v_i} = 10$ (resp.~$B_{{v}_i} = 11$).
Next we suppose $u_i = v_i$.
If $(\sigma_{2i}, \sigma_{2i-1}) = (+,+), (-,+), (-,-)$, then we set
$B_{u_i} =B_{v_i} =00, 10, 11$ respectively.
By putting the blocks $\mathcal{B}_{01}$ and $\mathcal{B}_{10}$
appropriately on the remaining positions for blocks, we can retrieve the 
original path $p \in \mathcal{P}_{\Lambda_0}$.

Next we consider the case of $a=1$.
Since every $p \in \mathcal{P}_{\Lambda_1}$ satisfies the relation 
$\varphi_1(p) - \varepsilon_1(p) = 1$, the number of up spinons is larger than
that of down spinons by exactly one in $p$.
Noting this difference,
the remaining arguments go almost in the same way as in the case of $a=0$, {so we omit their details. }

%
%

\bibliography{BIB}
\end{document}